\documentstyle[aps,epsf]{revtex}
\begin{document}
\draft
\twocolumn[\hsize\textwidth\columnwidth\hsize\csname
@twocolumnfalse\endcsname
\title{Superconducting vortex avalanches, voltage bursts, and vortex
plastic flow: effect of the microscopic pinning landscape on 
the macroscopic properties}
\author{C. J. Olson, C. Reichhardt, and Franco Nori}
\address
{Department of Physics, The University of Michigan, Ann
Arbor, Michigan 48109-1120}
\date{\today}
\maketitle
\begin{abstract}
Using large scale simulations
on parallel processors,
we 
analyze in detail the dynamical behavior of superconducting 
vortices undergoing avalanches.
In particular, we quantify the effect of the pinning landscape on the 
macroscopic properties of vortex avalanches and vortex plastic flow.
These dynamical instabilities are triggered
when the external magnetic field is increased
slightly, and are thus driven by a flux gradient 
rather than by thermal effects.
The flux profiles, composed of rigid flux lines that interact with 100 or 
more vortices,
are maintained in the Bean critical state and do not decay away from it.
By directly determining vortex positions during avalanches
in the plastically moving lattice,
we find that experimentally observable voltage bursts correspond
to the pulsing movement of vortices along branched channels or winding chains
in a manner reminiscent of lightning strikes.
This kind of motion cannot be described by elastic theories.
We relate the velocity field and cumulative patterns of vortex
flow channels with statistical
quantities, such as distributions of avalanche sizes. 
Samples with a high density of strong pinning sites produce 
very broad avalanche distributions.
Easy-flow vortex channels appear in samples with a low pinning density,
and typical avalanche sizes emerge in an otherwise broad 
distribution of sizes.
We observe a crossover 
from interstitial motion in narrow channels to pin-to-pin motion in
broad channels as pin density is increased.
\end{abstract}
\pacs{PACS numbers: 64.60.Ht, 74.60.Ge}
\vskip2pc]
\narrowtext

\section{Introduction}

Dissipative extended systems that are slowly driven 
to marginally stable states produce complex and novel dynamics.
Systems characterized by {\it avalanche} dynamics, in which energy dissipation
occurs in sudden bursts of collective activity, are of particular interest.
Avalanche behavior has been studied in many systems,
including granular assemblies \cite{1,2,3,4,5,6},
magnetic domains \cite{7}, charge density waves \cite{8},
fluid flow down inclines \cite{9,10},
and flux lines in type-II superconductors \cite{11,12,13}.

Microscopic information can reveal why some regimes of material 
parameters produce broad distributions of avalanche sizes,
while others do not.  Such information
is generally unavailable, however,
and only limited macroscopic changes in system configurations
can be observed.  
Further, experimentally tuning microscopic parameters and recording
detailed {\it microscopic} information about the dynamics
is difficult.
Numerical simulations allow exact control of microscopic
parameters, such as pin strength and density, 
and provide both very precise microscopic information, such 
as individual vortex motion, and macroscopic information, such as voltage
signals. 
Comparison of simulation and experiment has been hampered 
because the extreme numerical demands of accurate models
has caused virtually all theoretical explorations
of these systems (see, e.g., Ref.\cite{14})  
to employ discrete dynamics governed by 
simple rules.
With recent advances in parallel processing, however, it 
is now possible to study much more realistic and complicated
models using molecular dynamics (MD) simulations.
In this paper, we present a continuous, MD simulation
of superconducting samples containing a critical state of
very slowly driven vortex lines that undergo avalanching behavior.
We find that the density of pinning sites plays an important role in
producing or suppressing broad distributions of avalanches.
When the pin density is low, favoring easy-flow vortex channels,
characteristic avalanche sizes appear.  For higher densities,
no unique channels form, and distributions remain very broad.

Flux penetrates a type-II superconductor in the form of discrete
quantized vortices that repel each other and are attracted by
defects in the superconducting material.  A gradient in the vortex density
develops and drives vortices into the material. A balancing
pinning force holds the vortices in a metastable state,
known as the critical state or Bean state \cite{15}.  
As an external field is slowly increased,
additional flux lines enter the sample and occasionally
cause large disturbances.  Although experiments can use local changes
in flux to detect the motion of these vortices
\cite{11,13,16,17,18}, at present the
vortex motion cannot be directly imaged over long enough time scales
to permit a statistical characterization of the motion.
Thus, computationally generated information on
vortex movements is of great interest.

To investigate dynamical instabilities producing cascades of flux
lines in superconductors, we have performed extensive
MD simulations on parallel multiprocessors using a wide variety
of relevant parameters that are difficult to {\it continuously\/}
tune experimentally, such as vortex density $n_v$, pinning density $n_p$,
and maximum pinning force (or strength) $f_p$.
We do not observe a parameter-independent universal
response to perturbations in our samples,
but instead find a rich variety of behaviors
in which all of these parameters play an important role.
This ranges from the collective motion of vortex chains
dominated by pin-to-pin transport to 
the appearance of very narrow interstitial channels, 
where vortices flow between pinning sites,
in agreement with recent Lorentz microscopy results \cite{16}.
Computer simulations 
\cite{19,20,21} are a
valuable tool for the analysis of the {\it microscopic spatio-temporal
dynamics} of individual flux lines in superconductors and its
relation with commonly measured macroscopic averages.
With simulations, interactions between a plastic vortex lattice and
rigid pinning sites can be easily examined \cite{22}.

We use a $T=0$ MD algorithm to perform 
large-scale, detailed simulations of
many superconducting samples.
Each sample has one of three pinning strengths $f_{p}$
and one of three pinning densities $n_{p}$.  Five different
combinations of $f_{p}$ and $n_{p}$, spanning a wide variety of possible
pinning configurations, were considered in this work.
We study {\it dynamical}, rather than {\it thermal}, instabilities; thus,
thermally-activated flux creep \cite{23}
does not occur in our system, 
and all avalanches are driven solely by the competition between the
vortex gradient and pinning forces.
Each avalanche is triggered by the addition of a {\it single} flux line to 
the system.  After the avalanche ends and the system 
reaches mechanical equilibrium, another flux line is added to the system.
That is, during each avalanche, {\it no} vortices are added to 
the sample.
The work presented here is distinct from previous MD simulations
in several ways.  
Our detailed study involves
more than $10^4$ avalanches for each combination of $f_{p}$ and $n_{p}$,
recorded using an extremely large number of MD time steps 
($10^{4}$ hours on an IBM SP parallel computer).  This
allows us to construct reliable, statistically significant
distributions from the avalanches. 
Our more realistic two-dimensional model also employs 
a much longer vortex-vortex interaction range than previously
used \cite{20}.

This paper is organized as follows.  In section II, a detailed 
description of the numerical simulation is given.
Section III contains our definition of a vortex avalanche.
In section IV, we present a variety of images of the vortex
avalanche events, highlighting common features
such as vortex motion in winding chains
and the pulse-like nature of the avalanches.
In section V, statistical distributions are
analyzed. Features in these distributions are related quantitatively to
the microscopic parameters and dynamics of the system.  
Sections VI and VII contain 
brief comparisons of our work to recent experiments.
Finally, we summarize our results in Section VIII.

\section{Simulation}

We model a transverse two-dimensional slice 
(in the $x$--$y$ plane) of an infinite
zero-field-cooled superconducting slab containing
rigid vortices that are parallel
to the sample edge (${\bf H}=H{\hat {\bf z}}$).
The sample is periodic in the $y$ direction only, and there
are no demagnetizing effects.  
An external field is modeled by the presence of flux lines in an
unpinned region along one edge of the sample.  This field is very slowly
increased by adding a single vortex to the unpinned region
each time the sample reaches a state of mechanical equilibrium.
Vortices enter the superconducting slab under the force of
their mutual repulsion and pass through
a pinned region $24\lambda \times 26\lambda$ in size, where $\lambda$ is
the penetration depth.
They form a flux gradient naturally due to their own interactions 
\cite{19} and
give rise to the critical current $dB/dx= 2\pi J_{c}/c$. 
Experiments employ a flux profile which, on average, does not change with time
inside the sample \cite{11}.  To model this, we remove 
vortices from the simulation when they
exit the opposite end (i.e., the right edge in our figures)
of the pinned region.  

The vortex-vortex repulsion is correctly represented
by a modified Bessel function, $K_{1}(r/\lambda)$, cut off beyond
$r=6 \lambda$ where the force is quite small \cite{24}.  
The vortices also interact with a large number of non-overlapping
short-range attractive parabolic wells of radius $\xi_{p}=0.15\lambda$.
The maximum pinning force, $f_{p}$, of each well in a given sample 
is uniformly distributed between $f_{p}^{\rm max}/5$ and $f_{p}^{\rm max}$.
The pinning strength in our samples varies
from very weak pins ($f_{p} \lesssim 0.3 f_{0}$) to very strong pins 
($f_{p} \gtrsim 2.0 f_{0}$), with maximum pinning values
given by $f_{p}^{\rm max}=0.3f_{0}$, $1.0f_{0}$, or $3.0f_{0}$.
Here, 
$$f_{0} = \frac{\Phi_{0}^{2}}{8\pi^{2}\lambda^{3}} \ .$$
The samples studied here contain one of three different
densities of pinning sites, $n_{p}$, corresponding to three different
numbers of pinning sites, $N_{p}$: $n_{p}=5.93/\lambda^{2}$ ($N_{p}=3700$),
$n_{p}=2.40/\lambda^{2}$ ($N_{p}=1500$),
or $n_{p}=0.96/\lambda^{2}$ ($N_{p}=600$).
Although the pinning potential is fixed in time, the energy landscape
produced by the moving vortices evolves continuously.

The overdamped equation
of vortex motion is 
\begin{eqnarray}
{\bf f}_{i}={\bf f}_{i}^{vv} + {\bf f}_{i}^{vp}=\eta{\bf v}_{i} \ ,
\end{eqnarray}
where the total force ${\bf f}_{i}$ on vortex $i$ (due to other vortices 
${\bf f}_{i}^{vv}$, and pinning sites ${\bf f}_{i}^{vp}$) is 
given by
\begin{eqnarray}
{\bf f}_{i} &=& \ \sum_{j=1}^{N_{v}}\, f_{0} \,\ K_{1} \hspace{-2pt}
\left( \frac{ |{\bf r}_{i} - {\bf r}_{j} | }{\lambda} \right)
\, {\bf {\hat r}}_{ij} \nonumber
         \\ & + & \sum_{k=1}^{N_{p}} \frac{f_{p}}{\xi_{p}} \
|{\bf r}_{i} - {\bf r}_{k}^{(p)}| \ \ \Theta \hspace{-2pt} \left(
\frac{ \xi_{p} - |{\bf r}_{i} - {\bf r}_{k}^{(p)} |}{\lambda} \right) \
{\bf {\hat r}}_{ik} \, .
\end{eqnarray}
Here, $ \Theta$ is the Heaviside step function, 
${\bf r}_{i}$ is the location of the $i$th vortex,
${\bf v}_{i}$ is the velocity of the $i$th vortex, 
${\bf r}_{k}^{(p)}$ is the location of the $k$th pinning site, 
$\xi_{p}$ is the radius of the pinning site,
$N_{p}$ is the number of pinning sites,
$N_{v}$ is the number of vortices,
${\bf {\hat r}}_{ij}=({\bf r}_{i}-{\bf r}_{j} )/
|{\bf r}_{i} -{\bf r}_{j}|$,
${\bf {\hat r}}_{ik}=({\bf r}_{i}-{\bf r}_{k}^{(p)})/
|{\bf r}_{i}-{\bf r}^{(p)}_{k}|$,
and we take $\eta=1$. 
We measure all forces in units of
$f_{0}=\Phi_{0}^{2}/8\pi^{2}\lambda^{3}$,
magnetic fields in units of $\Phi_{0}/\lambda^{2}$,
and lengths in units of the penetration depth $\lambda$.
The number of vortices forming the Bean state varies from sample
to sample, ranging from a minimum value of 
$N_{v} \approx 240$ for the sample with
a low density of strong pins, 
$n_{p}=0.96/\lambda^{2}$ and $f_{p}^{\rm max}=3.0f_{0}$, to
a maximum value of 
$N_{v} \approx 1700$ for the sample with 
a high density of strong pins, $n_{p}=5.93/\lambda^{2}$
and $f_{p}^{\rm max}=3.0f_{0}$. 
Throughout this paper we examine the regime 
$0.06 \lesssim n_{v}/n_{p} \lesssim 0.82$, where $n_{v}$ is the vortex density.

\begin{figure}
\centerline{
\epsfxsize=3.5 in
\epsfbox{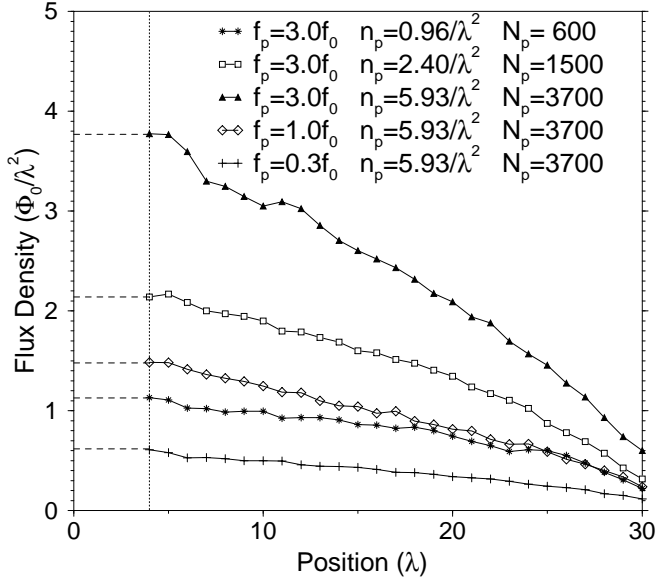}}
\vspace{0.15in}
\caption{Magnetic flux density profiles $B(x)$ for each sample studied.
The profiles were obtained using
$B(x)=(24\lambda)^{-1} \int_{0}^{24\lambda} d\!y \, B(x,y)$.
The area 
$0<x<4 \lambda$, $0<y<24\lambda$,
to the left of the dotted line, 
is the unpinned region through which vortices enter the
sample.  This mimics the external field; the 
dashed line indicates the field strength in this region.
The presence of bulk pinning in the sample (located in the region
$4\lambda <x<30 \lambda$, $0<y<24\lambda$)
provides a barrier for flux entry and exit
at the interface $x=4\lambda$.
The five profiles correspond to five samples with three different 
densities of pinning sites, $n_{p}$, 
and three different uniformly distributed pinning strengths,~$f_{p}$.
Starting from the top (very strong pinning) to the bottom, we have
filled triangles: $f_{p}^{\rm max}=3.0 f_{0}$, 
$n_{p}=5.93/\lambda^{2}$, $N_{p}=3700$, $N_{v} \approx 1700$;
open squares: $f_{p}^{\rm max}=3.0 f_{0}$,
$n_{p}=2.40/\lambda^{2}$, $N_{p}=1500$, $N_{v} \approx 1000$;
open diamonds: $f_{p}^{\rm max}=1.0 f_{0}$,
$n_{p}=5.93/\lambda^{2}$, $N_{p}=3700$, $N_{v} \approx 700$;
asterisks: $f_{p}^{\rm max}=3.0 f_{0}$, 
$n_{p}=0.96/\lambda^{2}$, $N_{p}=600$, $N_{v} \approx 500$; and
plus signs: $f_{p}^{\rm max}=0.3 f_{0}$, 
$n_{p}=5.93/\lambda^{2}$, $N_{p}=3700$, $N_{v} \approx 240$.
Note that the slope of $B(x)$, i.e., $J_{c}(x)$, is somewhat larger
towards the right edge of the sample where the flux density is lower and
the effective pinning is larger.
The average slope is not altered by avalanches since the majority
of the vortices in the sample do not move during an avalanche.
}
\label{fig:1}
\end{figure}

\section{Vortex Avalanches}

The flux profile in the sample (see Fig.~\ref{fig:1}) is maintained in a 
{\it quasi-magnetostatic state}, with only
a single vortex added to the left side of the sample every time
the sample reaches mechanical equilibrium \cite{23}.
Although this resembles granular sandpile experiments that add a
single grain at a time to the apex of the pile \cite{1,2},
there are very important differences between granular
systems and our vortex system:  the
vortex interaction has a much longer range, incorporating the effects
of up to 100 nearest neighbors;  inertial effects are negligible
in this highly overdamped system;  and the disorder is quenched (i.e., pinning
sites are fixed) and tunable (that is, the effective pinning strength can
be varied with field).

\begin{figure}
\centerline{
\epsfxsize=6.4 in
\epsfbox{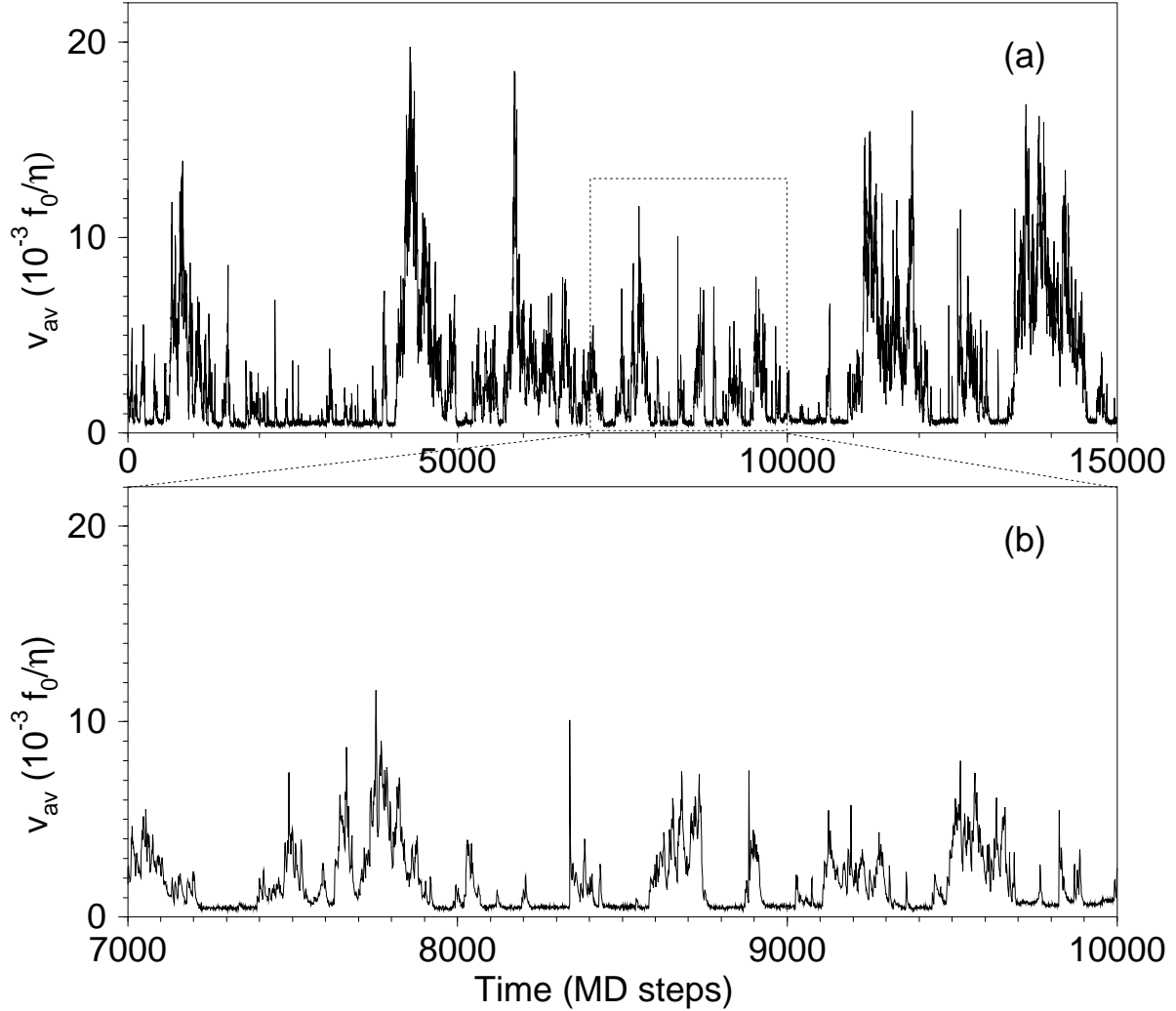}}
\caption{Plot of the average vortex velocity $v_{\rm av}$,
shown in normal (a) and
expanded (b) views.  The box in (a) indicates the region expanded
in (b).
This quantity resembles voltage signals induced by moving
vortices in actual experimental samples [11]. 
An additional vortex is added to the left edge of the sample only when 
$v_{\rm av}$ falls below the threshold (background) value of 
$v_{\rm th}=0.0006 f_{0}/\eta$.
Every period of avalanche activity can be clearly resolved.  Certain
velocity bursts composed of several closely spaced peaks in $v_{\rm av}$
correspond to single avalanches in which
{\it several} vortex chains move consecutively in a manner 
resembling lightning strikes.
In these avalanches with multiple peaks in $v_{\rm av}$, the disturbance
does not propagate through the sample at constant $v_{\rm av}$ in one
pulse, but moves in repeated pulsing waves.
This signal is from a sample containing 1700 vortices
interacting with a high density of  strong
pinning sites: $n_{p}=5.93/\lambda^{2}$, 
$f_{p}^{\rm max}=3.0f_{0}$.
}
\label{fig:2}
\end{figure}

To gauge the amount of vortex motion occurring in a sample at each
MD time step, we use the average 
vortex velocity $v_{\rm av}$,
given by the sum of
the magnitude of the velocity of each vortex
$v_{i}$ divided by the number of vortices in the 
sample $N_{v}$,
\begin{eqnarray}
v_{\rm av}=\frac{1}{N_{v}} \ \sum_{i=1}^{N_{v}} v_{i} \ .
\end{eqnarray}
We add a new vortex only when significant motion has stopped,
as indicated when 
\begin{eqnarray}
v_{\rm av} < v_{\rm th} \ ,
\end{eqnarray}
where $v_{\rm th}$ is a low threshold value. 
The signal $v_{\rm av}$ is plotted in Fig.~\ref{fig:2}
for a sample with a high density of
strong pinning sites and a threshold level $v_{\rm th}=0.0006 f_{0}/\eta$. 
Each time a new vortex is added to the sample,
a very small peak appears in $v_{\rm av}$.
Larger peaks are produced during the avalanches that are occasionally triggered
by the addition of a vortex, with the largest peaks
generated by a combination of either
a few vortices moving rapidly or
a large number of vortices moving slowly.
Many avalanche events produce $v_{\rm av}$ 
signals that consist of several peaks clustered together, 
indicating that the avalanche disturbance does {\it not}
propagate through the sample at constant $v_{\rm av}$, 
but moves in {\it repeated} pulsing
waves \cite{25}.  
This burst-like behavior is a result of the combined effects of the
vortex density gradient and the two-dimensional nature of the vortex motion.  

Most avalanche disturbances start at the outer sample edge 
(left edge in the figures)
where the vortex density is highest.
When a vortex inside the sample is displaced from its pinning well
due to a small increase in the external field, pinned vortices
ahead of it prevent it from moving down the gradient.
This is in contrast to granular sandpiles, where sand grains can move
in the third dimension to avoid such blockages.
A string of vortices is depinned in a domino-like effect,
with each moving only far enough to depin the next vortex and then stopping.
This pulsing motion continues until the forces on all vortices in
the sample are once again below the threshold depinning force.

\twocolumn[\hsize\textwidth\columnwidth\hsize\csname
@twocolumnfalse\endcsname
\vskip2pc]

By adding only one vortex at a time, we probe the system in the limit 
of a zero current of incoming vortices.  
Large values of the current do not produce well-defined individual avalanches,
so it is important that a given avalanche die away completely before
adding the 
next vortex.
Adding many vortices simultaneously, or at a very fast rate, can significantly
modify the state of the system.  The information we learn about a 
rapidly driven system tells us
not about the original state but about the 
(possibly drastically) modified state.
Use of an effectively 
infinitesimal perturbation allows us to avoid altering 
the nature of the original system while still gaining information about it,  
something that is important in other systems as well.
For instance, in classical mechanics, a very small perturbation  is applied
to a particle to test the stability of its orbit.  In electromagnetic
systems, a small test charge is added to systems in order to probe
their electrical screening or dielectric properties.
In many-body physics, a very small electric or magnetic field is applied
to systems to determine their diamagnetic response.

\subsection{Voltage bursts and intermittency}

The time-evolution of the velocity signal plotted in Fig.~\ref{fig:2} 
resembles that observed for a variety of different physical
systems in the so-called intermittent regime \cite{26,27}.
In particular, the general features of the velocity signal
in Fig.~\ref{fig:2} are similar to the fluid velocity 
at a point in the interior of the container
in Rayleigh-B\'{e}nard 
convection experiments \cite{28}
in the intermittent regime.  
In the vortex and fluid cases, the somewhat regular velocity signals are 
intermittently interrupted by velocity ``bursts'' of finite duration
which occur at seemingly random times.
This section explores the conceptual analogies and differences
between the velocity bursts produced by vortices moving in 
superconductors and the velocity bursts observed in intermittency.
A quantitative and detailed comparison is beyond the scope of this
section, and will be presented elsewhere.

Let us first summarize the intermittent transition to chaos
\cite{29}.  More information can be found in Refs.~\cite{26,27}.
Intermittency refers to a signal that alternates in an apparently
random manner between long regular phases, known as laminar
phases or intermissions, and relatively short irregular bursts,
called chaotic bursts.   The frequency of these bursts 
increases with an external parameter that we call $\epsilon$ here.
Thus, intermittency provides a continuous route from 
regular, burst-free motion, for $\epsilon=0$, 
to chaotic motion, for large enough $\epsilon$.  
For very small values of $\epsilon$, 
there are long stretches of time, called laminar phases,
during which the dynamics is regular, remaining very close to  
the burst-free $\epsilon = 0$ fixed point.

Let us now consider a marginally stable Bean state.
Take $\epsilon$ to be the rate at which vortices are added 
to the system (i.e., the driving rate).
If $\epsilon = 0$, the dynamical system of vortices is at 
a fixed point in both position and velocity spaces.  
In particular, its average velocity $v_{\rm av}$ is zero.
In the language of nonlinear dynamics, the laminar phase
now has an infinite duration and the motion is regular; that is,
it is a fixed point of zero velocity.

When vortices are added to the system at a very small rate $\epsilon$,
such as one vortex added every $10^5$ MD steps, 
some vortices in the system rearrange their positions.
In this case, the average vortex velocity 
remains very close to the $v_{\rm av}=0$ laminar phase value,
up to small oscillations produced by vortex rearrangements,
In other words, if $\epsilon > 0$ and very small, the 
velocity of the system remains quite near its $v_{\rm av}=0$ fixed point
(see, for instance, Fig.~50(b) of Ref.~\cite{26}).
This dynamics near the fixed point, or ``inside the laminar region'',
does not continue forever. Eventually, the system exhibits a burst;
that is, $v_{\rm av}(t)$ suddenly increases to a larger value, producing 
a crest of apparently random shape, or a ``chaotic burst''.
After a period of time that is typically short, this burst suddenly comes
to an end, and the system is ``reinjected into'' the laminar region,
exhibiting its usual dynamics near the fixed point.
When vortices are added to the system 
at an increasingly large rate $\epsilon$,
the average vortex velocity displays more frequent bursts
away from the $v_{\rm av}=0$ fixed-point laminar phase value.

For chaotic systems in the intermittent regime, 
the length of time the system spends around its fixed point, 
and the length of time the system spends in the chaotic region, 
are both unpredictable.  
If the same computer run is repeated with initial conditions that
vary slightly, 
the details of the dynamics will not be reproduced \cite{26,27}. 
These and many other results have been obtained for chaotic 
systems in the intermittent regime.  Analogous calculations
have not been made for vortices in the Bean state.
It would be interesting to calculate, among other things, 
the average time the system spends around its fixed point 
(i.e., in the laminar region) as a function of the driving 
rate $\epsilon$.  It is unclear at this point how these 
quantities compare for the vortex and standard nonlinear systems.
To compute these quantities for vortices is far more complicated
than for the usual systems.  Fluid flow, modeled by the three Lorentz
equations, and other intermittent systems
(e.g., nonlinear RLC circuits) can be described effectively by few degrees
of freedom.  The problem of vortex avalanches requires solving
very many degrees of freedom, which greatly complicates the
calculations.  These will be considered elsewhere.

\section{Analysis of the spatiotemporal evolution of vortex avalanches}

The spatiotemporal propagation of avalanches through a sample
are of great interest.  The analysis presented
below provides insight into
which microscopic properties of a sample are most important in
determining the nature of the resulting avalanches.
Few experimental studies in any avalanche system are able to view individual
avalanches.
Experiments involving vortex avalanches so far have been able to
resolve only
a macroscopic result of the avalanche, such as flux leaving the
sample, and have not been able to determine the path the vortices
followed as they moved through the sample.  
Our simulations can easily image vortex avalanches,
revealing how changing the microscopic pinning
parameters affects the nature of the avalanches.

\subsection{Spatial configurations of vortex avalanches}

We first determine the path an avalanche follows through 
the sample by identifying the vortices involved 
\twocolumn[\hsize\textwidth\columnwidth\hsize\csname
@twocolumnfalse\endcsname
\begin{figure}
\centerline{
\epsfxsize=6.6 in
\epsfbox{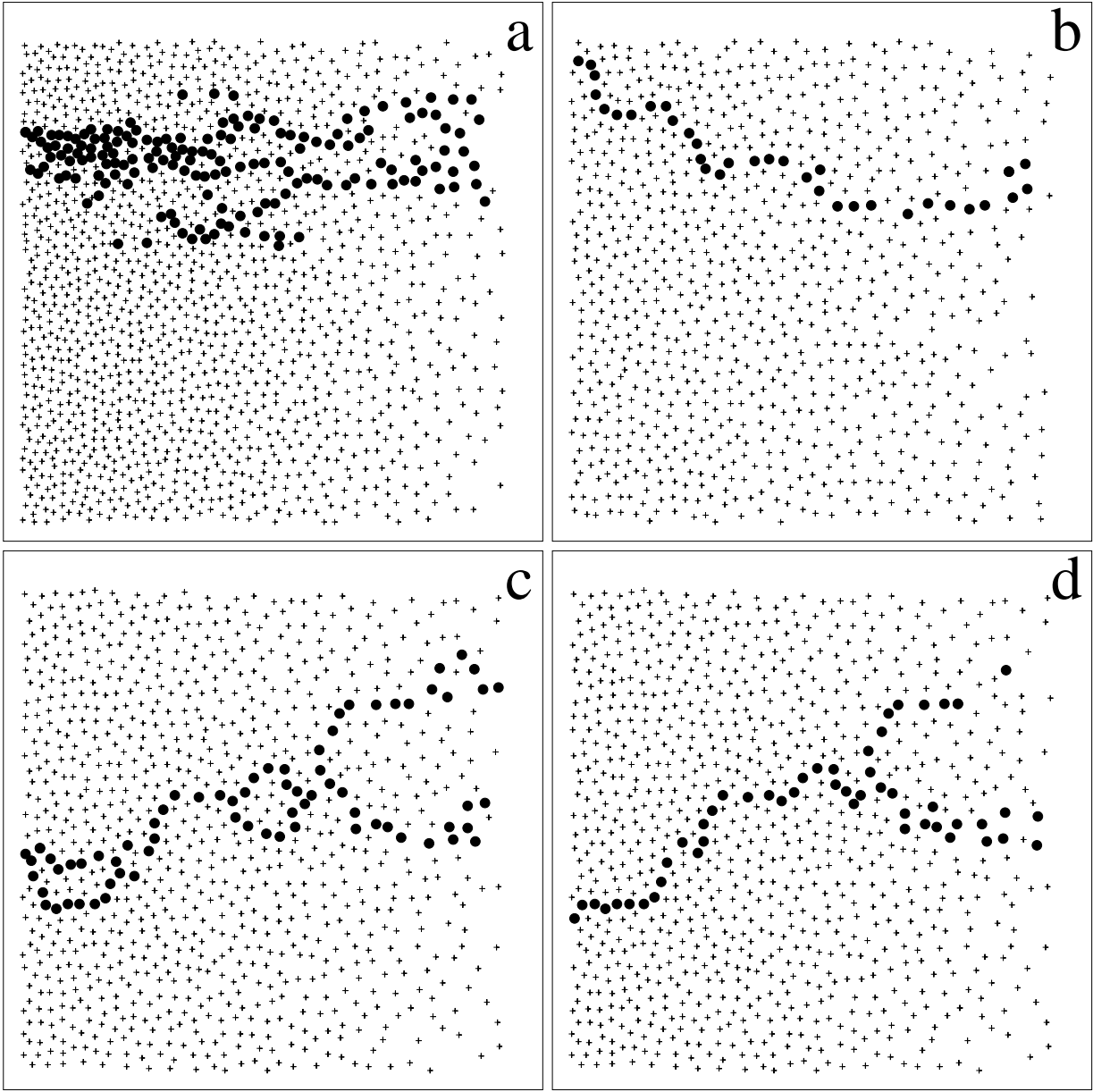}}
\caption{Snapshots of vortices participating in an avalanche event.
Here, the initial positions of vortices that
were depinned during the avalanche
are represented by filled circles.
Vortices that remained pinned are represented by crosses.
Vortex motion is towards the right of the figure, and
the entire $26 \lambda \times 24 \lambda$ sample 
is shown.  
Panel (a) is from a sample with $n_{p}=5.93/\lambda^{2}$ and
$f_{p}^{\rm max}=3.0f_{0}$;
panels (b), (c), and (d) are from a sample with lower 
pinning density: $n_{p}=2.40/\lambda^{2}$ and $f_{p}^{\rm max}=3.0 f_{0}$.
(a) shows a large event.  The stationary vortices in 
and around the flow path
are more strongly pinned than the moving vortices and provide barriers
to the flow.  
(b) is a typical chain-like event.  Most
events are this size or smaller.
(c--d) show how a characteristic channel for motion 
can change slightly over time
[e.g., from (c) to (d)].  Infrequent large events rearrange the
vortices in (c) and alter the interstitial 
pinning caused by vortex-vortex interactions,
resulting in the new channels shown in (d).
In other words, the channels in (d) are produced by
a vortex lattice rearrangement after the channels in (c) dominated
the transport over several avalanches.
Note the gradient in the vortex density, with a higher density on the
left side of the figure.
}
\label{fig:3}
\end{figure}
\vskip2pc]

\hspace{-13pt}
in representative avalanche events.
A vortex is considered an avalanche participant if it
is depinned, indicated when its 
displacement $d_{i}$ during the time interval between adding new
vortices to the sample
is greater than the pinning diameter 
$2 \xi_{p}$, 
\begin{eqnarray}
d_{i} > 2 \xi_{p} \ .
\end{eqnarray}
This distinction is necessary
since most vortices in the sample are displaced very slightly during
an avalanche, but only a small number depin and are transported down
the gradient.
Because our simulation operates in the regime
$n_{v} < n_{p}$, all vortices remain trapped at pinning sites 
when an avalanche is not occurring,
except in the case of very low pin density when some vortices
remain at interstitial sites created by the
repulsion from surrounding vortices.

A single avalanche event is depicted in
each panel of Fig.~\ref{fig:3}. The entire sample is shown
in each case, with 
the initial positions of vortices that were depinned during the
avalanche
marked with filled circles. 
The vortices that remained pinned are indicated with crosses.
Rare large events, such as that illustrated in Fig.~\ref{fig:3}(a)
for a densely pinned sample, involve a broad region of the sample.  
These events occur only after
a considerable amount of strain has accumulated in the vortex lattice.
In the most frequently occurring events, winding chains of vortices 
move from pinning site to pinning site, with each vortex moving into the
site vacated by a vortex to its right, as in Fig.~\ref{fig:3}(b).
The chains extend down the flux gradient in the $x$--direction, and
are not perfectly straight but wind in the $y$--direction.
The amount of winding increases as the pinning strength $f_{p}$ decreases.
The chains do not appear at the same location
during every avalanche, and chain size may vary:
in some events a chain spans the sample, while in others the chain contains
only three or four vortices.  

When the pinning density is high, moving vortex chains are equally likely
to form anywhere in the sample.
As the pinning density is lowered, however,
the probability that the movement will occur in a certain well-defined channel
at one location in the sample becomes very high.
The position of this channel varies from sample to sample, but all samples
with low pinning density contain such a channel.
An example of  a ``preferred channel'' 
is
illustrated in Fig.~\ref{fig:3}(c,d).
Here the vortices follow the {\it same} winding path
during several consecutive avalanches.  These meandering vortex
channels display 
{\it branching} behavior or form small loops around stronger
pinning sites.  As a result of the low pinning density in the sample shown in 
Fig.~\ref{fig:3}(c,d), {\it interstitial
pinning} is highly important.  
Interstitial pinning occurs when a vortex
is held in place only by the repulsion from surrounding core-pinned vortices. 
Since the strength of the interstitial pinning is significantly weaker 
than that of the surrounding
core pinning sites in a sample, as in Fig.~\ref{fig:3}(c,d),
the resulting avalanches follow easy-flow paths composed almost
entirely of interstitially pinned vortices.
The occasional large events in such a sample rearrange
the interstitial pinning landscape, 
leading to the formation of a new set of
flow channels that will persist for a period of time until the next
large event occurs.  
For example, the persistent flow channel of Fig.~\ref{fig:3}(c) 
was altered into the channel of Fig.~\ref{fig:3}(d) by a large event.

A small packet of flux never actually moves
from one end of the sample to the other in a single avalanche,
regardless of whether vortices move in fixed channels or in constantly changing
paths.
Instead, movement is transmitted from vortex to vortex, with an individual
vortex rarely moving more than one to two pinning sites away from its
former location during an avalanche.  Thus, the disturbance crosses the
sample, but a vortex does not, and the time span of a typical
avalanche, $\tau$, 
(measured below) is much shorter than the time required for a
single vortex to traverse the sample.  

\subsection{Snapshots of the velocity field}

The images analyzed above show that avalanche
disturbances involve several possible types of moving bundles
or chains of vortices,
but give no information about the time scales for
vortex motion.
Thus, to examine the avalanche dynamics, we consider the velocities of
individual vortices at several instants during an avalanche and
present a series of snapshots of the vortex movements during a 
{\it single} avalanche.  We find 
that avalanche disturbances propagate neither 
instantly nor at constant speed through the sample.

Figure~\ref{fig:4} shows an avalanche moving through a portion of
a sample with a high density of strong pinning sites.
An arrow of a length proportional to the instantaneous vortex velocity
marks the initial position of each moving vortex.
The velocity of the vortices near
the right edge of the sample is greater than that near the left edge due
to the vortex density gradient.
The uniform time interval separating each panel was chosen as
$t_{h}$, the time required for a vortex to hop from one pinning
site to the next,  so that most vortices complete
their motion within one panel of the figure.
The disturbance in this avalanche propagates
through the sample, with most of the vortices involved stopping
exactly {\it one} pinning site to the right of their initial positions.
Thus, the majority of the motion occurs inside the sample and cannot
be detected by probing only vortices exiting the sample.
Those vortices that do exit
have initial positions within one to two lattice units from the sample edge,
and we do not observe vortices moving many pinning sites to the 
right during a single avalanche in order to leave the sample. 
We therefore find that during the largest avalanches, when the greatest
number of vortices exit the sample,
all movement and flux exit occurs in a {\it wide} region of the sample.
Fig.~\ref{fig:3}(a) shows an example of a wide avalanche.

The event pictured in Fig.~\ref{fig:4} 
has a medium-length lifetime.  Events with longer lifetimes
often consist of {\it more 

\twocolumn[\hsize\textwidth\columnwidth\hsize\csname
@twocolumnfalse\endcsname
\begin{figure}
\centerline{
\epsfxsize=6.4 in
\epsfbox{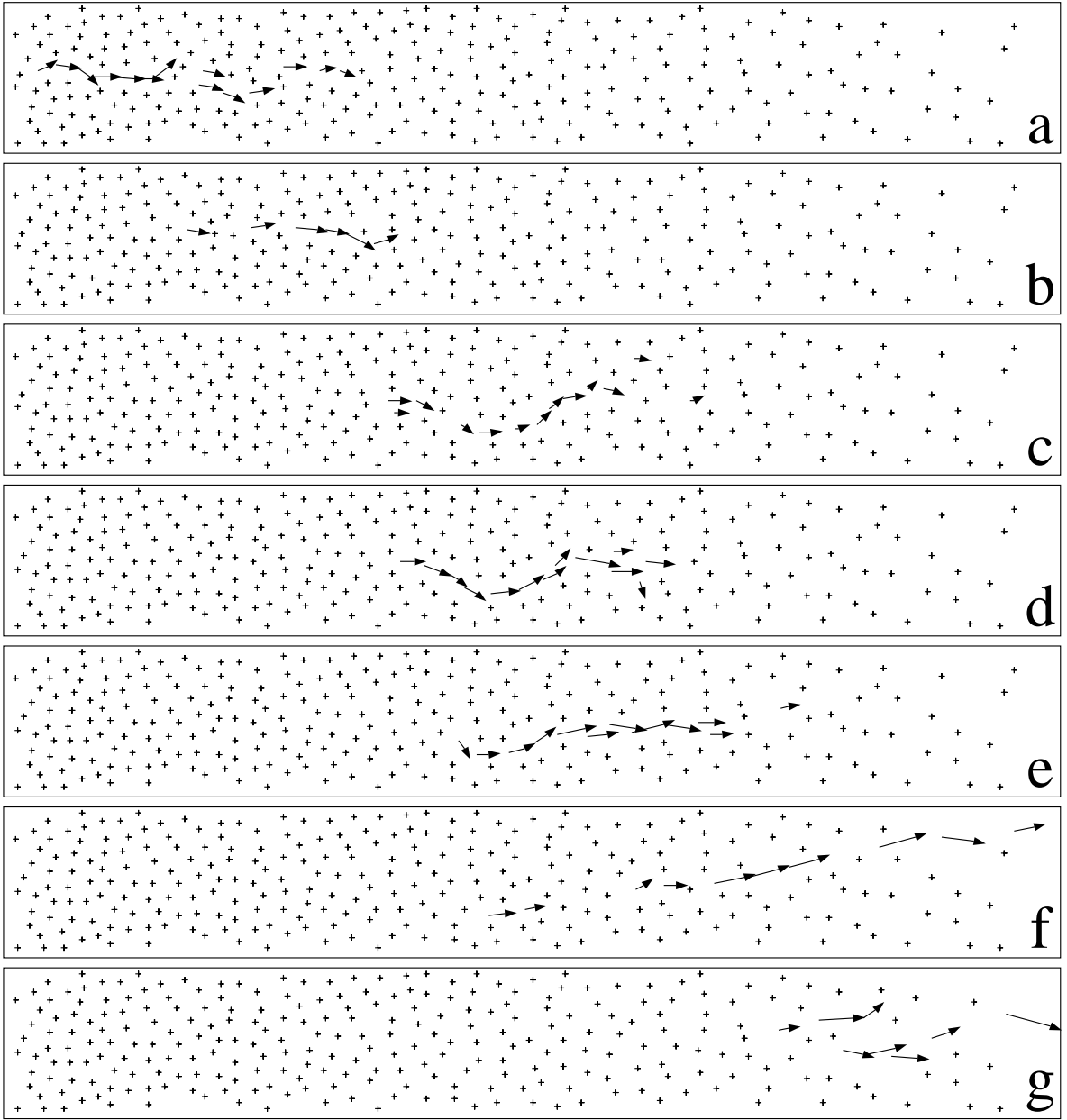}}
\vspace{0.15in}
\caption{
Consecutive snapshots of the vortex velocity field in a $25 \lambda \times
5 \lambda$ region
of a $26\lambda \times 24\lambda$
sample with a high density of  strong pinning sites,
$n_{p}=5.93/\lambda^{2}, f_{p}^{\rm max}=3.0 f_{0}$.
Each moving vortex is indicated by an arrow whose length is scaled by the
velocity of the vortex;  the remaining vortices are indicated by small crosses.
All vortices sit in pinning sites (not shown)
when not in motion.
A vortex is considered ``moving'' if it is depinned.
The remaining vortices are not completely motionless, but shift
very slightly inside
the pinning sites.
The disturbance propagates from the dense
left edge of the sample to the relatively less dense right edge.
The vortices in the rest of the sample (not shown) were not depinned.
The time interval between snapshots
is the typical time $t_{h}$ required for a vortex to hop from
one pinning site to another.
The illustrated motion is typical of a medium-sized avalanche in this sample.
}
\label{fig:4}
\end{figure}
\vskip2pc]

\twocolumn[\hsize\textwidth\columnwidth\hsize\csname
@twocolumnfalse\endcsname
\begin{figure}
\centerline{
\epsfxsize=6.4 in
\epsfbox{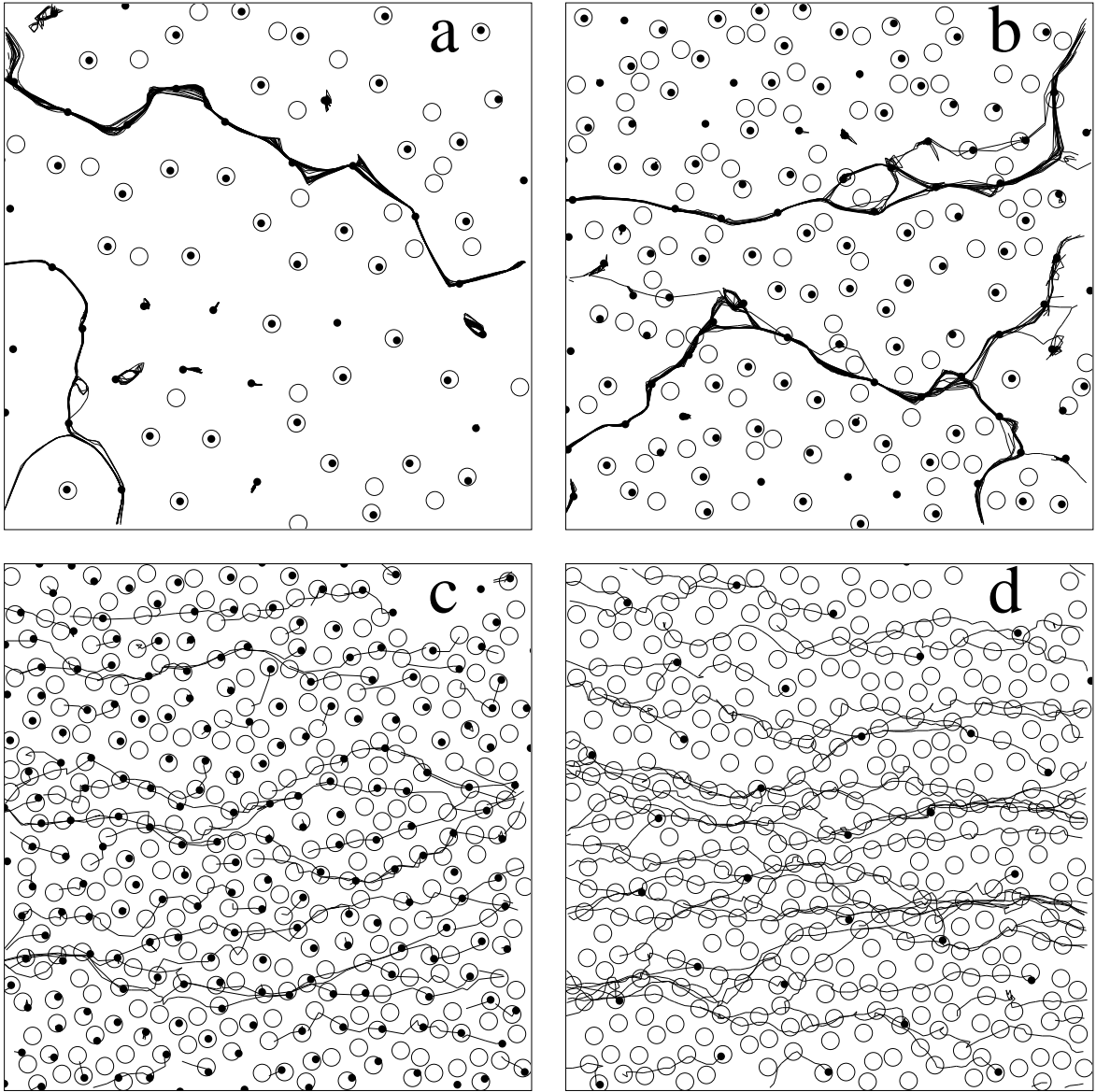}}
\vspace{0.15 in}
\caption{
Continuous lines indicate the paths vortices follow through a portion 
of the sample over an extended period of time covering many avalanches.
Vortices enter through the left edge and move towards the
right edge.  
Dots represent vortices, and open circles
represent pinning sites.  
An $8\lambda \times 8\lambda$
region of the $26\lambda \times 24\lambda$ sample is shown.  
(a) $n_{p}=0.96/\lambda^{2}$, $f_{p}^{\rm max}=3.0 f_{0}$.  
(b) $n_{p}=2.40/\lambda^{2}$, $f_{p}^{\rm max}=3.0 f_{0}$.
(c) $n_{p}=5.93/\lambda^{2}$, 
$f_{p}^{\rm max}=3.0 f_{0}$.
(d) $n_{p}=5.93/\lambda^{2}$, 
$f_{p}^{\rm max}=0.3 f_{0}$.
Vortex motion in the sample with $n_{p}=5.93/\lambda^{2}, 
f_{p}^{\rm max}=1.0 f_{0}$ 
resembles (c) and (d).  
In (a) and (b), strong pinning occasionally causes segments
of the vortex path to run towards the top and bottom of the
figure, transverse to the flux gradient.
The presence or absence of easy-flow channels
is strongly dependent on pin density.  
The channels present in (a) and (b) lead to avalanches with characteristic
sizes $N_{a}$ and lifetimes $\tau$.
Samples with higher pinning density do not have isolated easy-flow channels
and produce very broad distributions of
avalanche sizes.
Because of the higher pinning strength, avalanches 
in (c) have higher typical vortex velocities $v_{i}$ and shorter 
typical event lifetimes $\tau$ than those in (d),
resulting in tighter, less wandering vortex paths.}
\label{fig:5}
\end{figure}
\vskip2pc]

\hspace{-13pt}
than one pulse} of motion; that is, the average
velocity $v_{\rm av}$ exhibits several spikes or
oscillations during a single avalanche.

\subsection{Cumulative pattern of vortex flow channels}

The spatial configurations of the avalanches we observe are not
strongly affected by the sample pinning parameters, with all samples
producing vortex motion in winding chains that
propagate through the sample during one or more pulses.
The pinning parameters are important, however, when we consider
whether chains moving in consecutive avalanches
are concentrated in one area of the sample,
or whether they are evenly distributed throughout the sample over time.
By plotting the vortex trajectories with lines, and drawing these
lines for an extended period of time covering many avalanches,
we can identify the cumulative pattern of vortex flow channels for
different pinning parameters. 
Heavily-traveled regions of the sample are easily distinguished by a
concentration of trajectory lines.
An examination of small regions of samples shows that the typical
flow channel pattern varies with pinning density.

As shown in Fig.~\ref{fig:5}(a--b),
{\it interstitial channels} of easy flow for vortices
develop in samples with a low density of strong pinning sites.
The flow of flux lines
in Fig.~\ref{fig:5}(a) involves {\it only} the much more
mobile interstitial vortices moving {\it plastically} around their strongly
pinned neighbors, indicated by the fact that
the vortex trajectories form narrow paths that never intersect 
the pinning sites.
Similar behavior has been directly observed recently by Matsuda {\it et al.} 
\cite{16}, where
interstitial, chain-like avalanche flow was seen around a
few strongly pinned vortices 
\cite{30}.
As the pin density is increased, pin-to-pin vortex motion
becomes more important and the interstitial
channels become less well defined 
[Fig~\ref{fig:5}(b)]
until at the highest pin densities,  $n_{p}=5.93/\lambda^{2}$,
interstitial flow is no longer observed, and
the vortices always move from pinning site to pinning site.
This is illustrated in Fig.~\ref{fig:5}(c--d) 
for a high density of pins of two different strengths.
Note the absence of distinct isolated flow channels.

\subsection{Derivation of the criteria for the presence of interstitial
vortex motion}

In the limit of strong pinning, we expect interstitial vortex 
motion to occur whenever a vortex moving 
between two pinned vortices exerts a maximum force $f^{vv}_{\rm midpoint}$ 
on each pinned
vortex that is less than the pinning force $f_{p}$.  
This occurs at the midpoint between the two pinned vortices.
If this condition is not met, one of the pinned vortices will be
depinned, and the interstitial vortex might be trapped by the vacated
pinning site.
Given the pinning density $n_{p}$
of a sample, we can calculate the minimum 
pinning strength $f^{(p)}_{\rm min}$ required
to allow interstitial motion.
Assume that two vortices are pinned in adjacent wells separated
by the average distance $d_{p}=n_{p}^{-1/2}$ between pinning sites.
If an interstitial vortex passes directly between these two
pinned vortices, it is a distance $d_{p}/2$ from each pinned vortex,
and exerts a force on each equal to
\begin{eqnarray}
f^{vv}_{\rm midpoint} = f_{0} \ K_{1}\!\left(\frac{d_{p}}{2\lambda}\right) \ .
\end{eqnarray}
For interstitial motion to occur, we require
\begin{eqnarray}
f^{(p)}_{\rm min} = f^{vv}_{\rm midpoint} \ . \nonumber 
\end{eqnarray}
Thus, 
\begin{eqnarray}
f^{(p)}_{\rm min} = f_{0} \  K_{1}\!\left(\frac{d_{p}}{2\lambda}\right) \nonumber
\end{eqnarray}
\begin{eqnarray}
f^{(p)}_{\rm min} = f_{0} \ K_{1}\!\left(\frac{1}{2\lambda\sqrt{n_{p}}}\right) \ .
\label{eqn:interstitial}
\end{eqnarray}
For each of the pinning densities used in our simulation,
we can determine the minimum pinning strength $f^{(p)}_{\rm min}$ that
would permit interstitial motion. In samples with a high density of pinning
sites, $n_{p}=5.93/\lambda^{2}$, as in Fig.~\ref{fig:5}(c,d), we find
$f^{(p)}_{\rm min} = 4.7f_{0}$.  Since the actual pinning forces
in these samples are $f_{p} \lesssim 3.0f_{0}$, we do not expect 
interstitial motion to occur, and the simulations confirm that only
pin-to-pin motion occurs in these samples.
Samples with a lower pinning density of $n_{p}=2.40/\lambda^{2}$,
as in Fig.~\ref{fig:5}(b), have
$f^{(p)}_{\rm min} = 2.8f_{0}$.  In this case, with $f_{p} \lesssim 3.0f_{0}$,
occasional interstitial motion may occur near the strongest
pins, and a small amount of interstitial motion is observed in the 
corresponding simulations.
In samples with the lowest pinning density of $n_{p}=0.96/\lambda^{2}$,
as in Fig.~\ref{fig:5}(a),
$f^{(p)}_{\rm min} = 1.6f_{0}$.  This condition is easily met by a 
large portion of
the pins, which have $f_{p} \lesssim 3.0f_{0}$, and exclusively 
interstitial vortex motion is expected and observed in the simulations.

\section{Statistical distributions characterizing vortex avalanches}

The strikingly different behaviors of the vortex avalanches 
are quantified here by identifying the dependence of the distributions of 
several quantities on pinning parameters.   These quantities include:
$\tau$, the avalanche lifetime; $N_{a}$, 
the number of vortices participating in
the avalanche; $N_{f}$, 
the number of vortices exiting the sample during an event; 
$d_{\rm tot}$, 
the total vortex displacement occurring in the avalanche; 
as well as $d_{i}$ and $v_{i}$, the individual vortex displacements
and velocities during an avalanche.

\subsection{Avalanche lifetime}

A frequently used characterization of an avalanche is its total lifetime,
$\tau$, or the interval of time during
which the avalanche occurs.  In our simulation, $\tau$ is equal
to the interval between perturbations of the system
by the addition of a new vortex.  Recall that here, the
Bean state is always driven in a quasimagnetostatic mode:
a flux line is added, vortex positions in the sample shift and
may produce an avalanche, and the next flux line
is added only after the vortex lattice reaches
mechanical equilibrium.  We expect the avalanche lifetime to depend
on both the pinning strength and the pinning density.

The pinning strength $f_{p}$ influences avalanche lifetimes by
determining the speed of vortex motion in the sample.
We find a relationship between
the pinning strength $f_{p}$ and
the speed $v_{i}$ of an individual vortex 
by considering the behavior of a
vortex at position ${\bf r}_{i}$ immediately 
after it is depinned from a well at position ${\bf r}_{k}^{(p)}$.  
If we assume that the vortex barely has enough energy to
escape the well, the total force on the vortex just before it depins
is close to zero:
\begin{eqnarray}
{\bf f}_{i}={\bf f}_{i}^{vv} + f_{p}{\bf \hat{r}}_{ik} \approx 0 \ ,
\end{eqnarray}
where 
\begin{eqnarray}
f_{p}=| {\bf f}_{i}^{vp}(|{\bf r}_{ik}|=\xi_{p}) |
\end{eqnarray}
is the maximum pinning force at the edge of the parabolic well, 
${\bf \hat{r}}_{ik}= {\bf r}_{ik}/|{\bf r}_{ik}| = ({\bf r}_{i} - {\bf r}_{k}^{(p)})/
|{\bf r}_{i} - {\bf r}_{k}^{(p)}|$,
and ${r}_{ik}$ is the distance between vortex $i$ and the center
of the $k$th parabolic well.
Thus,
\begin{eqnarray}
{\bf f}_{i}^{vv} \approx - f_{p}{\bf \hat{r}}_{ik} \ .
\end{eqnarray}
The pinning force has an abrupt cutoff at the pinning radius $\xi_{p}$,
so when the vortex moves off the pinning site, the 
vortex-pin force ${\bf f}_{i}^{vp}$ suddenly falls to zero while
the long-range
vortex-vortex force ${\bf f}_{i}^{vv}$ changes only negligibly.
The resulting force on the vortex is
\begin{eqnarray}
{\bf f}_{i}={\bf f}_{i}^{vv} \approx -f_{p}{\bf \hat{r}}_{ik} \ .
\end{eqnarray}
Thus, the vortex velocity is
\begin{equation}
v_{i}=v_{c}=\frac{|{\bf f}_{i}|}{\eta} \approx \frac{f_{p}}{\eta} \ ,
\label{eqn:vortex_velocity}
\end{equation}
where $v_{c}$ is a characteristic velocity associated with the pinning
strength.
Vortices in samples with stronger pinning move faster when depinned than
vortices in samples with weaker pinning.

\begin{figure}
\centerline{
\epsfxsize=3.5 in
\epsfbox{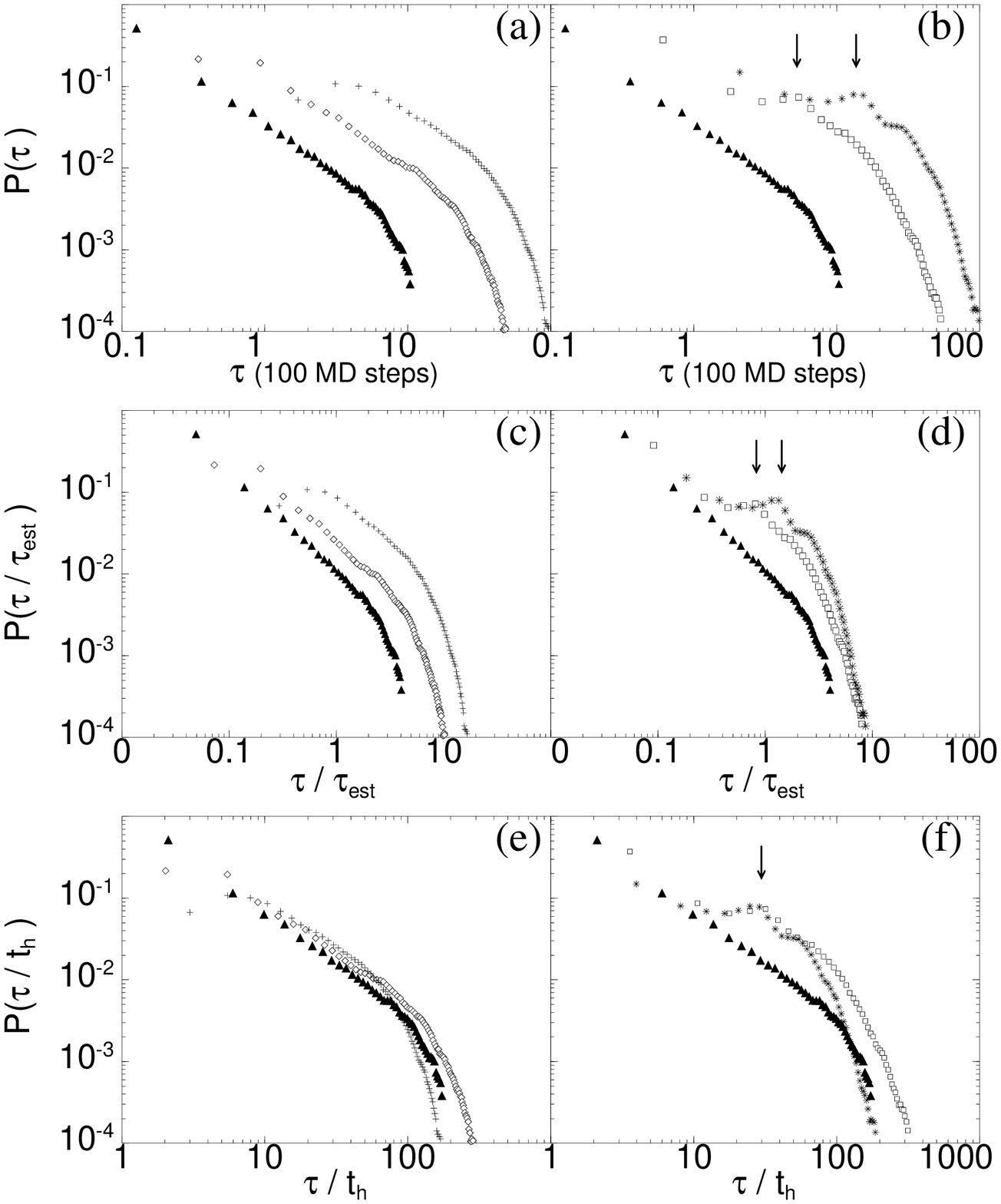}}
\centerline{
\epsfxsize=2.1 in
\epsfbox{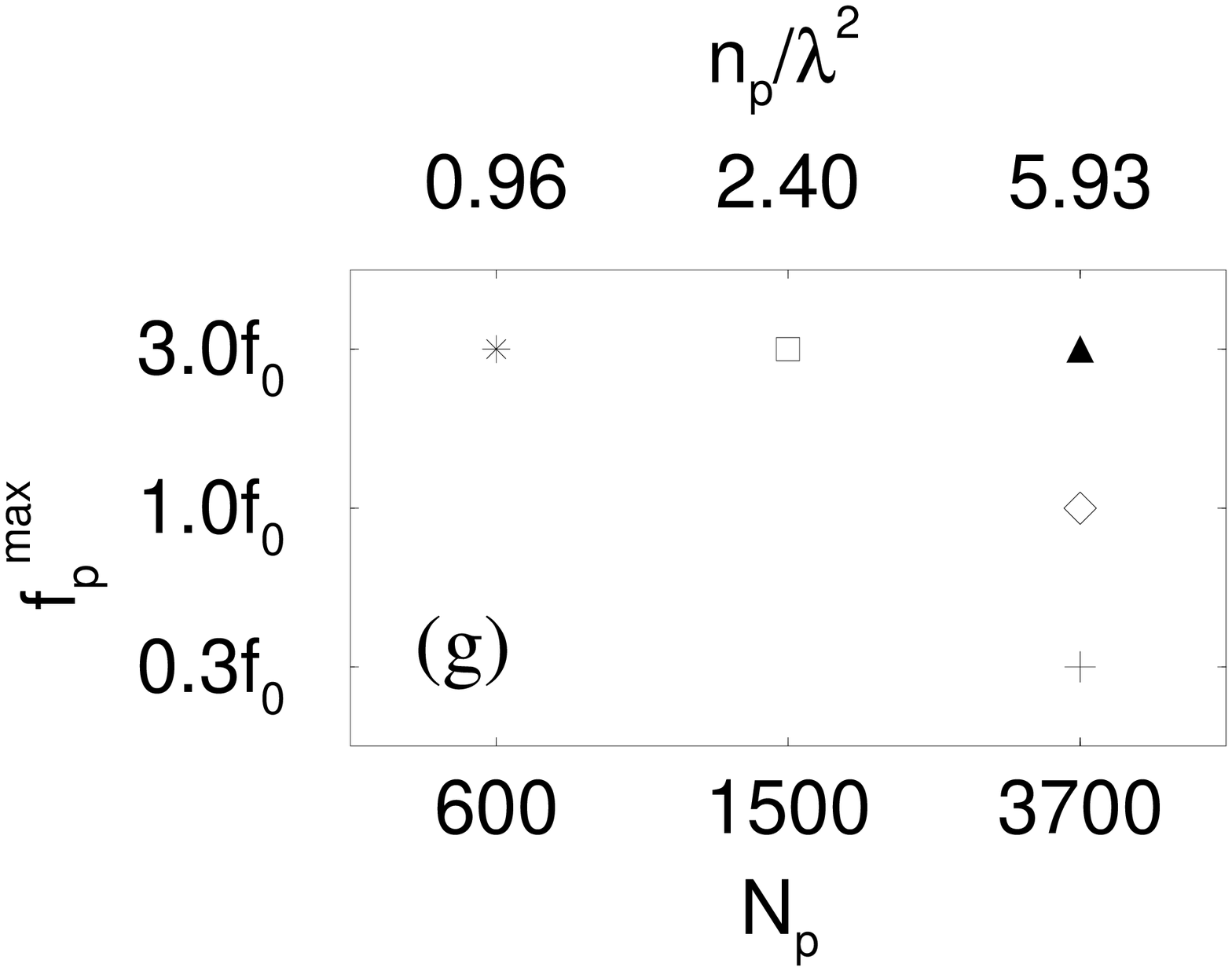}}
\caption{Distributions of avalanche lifetimes, $\tau$.
The leftmost panels (a,c,e) correspond to our highest density 
of pins, $n_{p}=5.93/\lambda^{2}$, 
and differing pinning strengths:
filled triangles, $f_{p}^{\rm max}=3.0 f_{0}$;  
open diamonds, $f_{p}^{\rm max}=1.0 f_{0}$;
plus signs, $f_{p}^{\rm max}=0.3 f_{0}$.
The right panels (b,d,f) correspond to samples with our strongest pins, 
$f_{p}^{\rm max}=3.0 f_{0}$,
and differing pinning densities:
filled triangles, $n_{p}=5.93/\lambda^{2}$;
open squares, $n_{p}=2.40/\lambda^{2}$;
asterisks, $n_{p}=0.96/\lambda^{2}$.
(c--d)~Distributions scaled by the estimated lifetime, $\tau_{\rm est}$.
Note the appearance of characteristic avalanche times 
indicated by arrows in (d)
at $\tau / \tau_{\rm est}\approx 0.8$ and $\tau / \tau_{\rm est}\approx 1.3$. 
(e--f) Distributions scaled by the hopping time, $t_{h}$.
In (e) the hopping time between pinning
sites is $t_{h}=$6, 18, and 57 MD steps, respectively.
Also in (e), notice the power law behavior 
$P(\tau/t_{h}) \sim (\tau/t_{h})^{-1.4}$
for small $\tau/t_{h}$ over two decades
for the sample with a high density of strong pins (filled triangles).
In (f) the hopping time between
pinning sites is $t_{h}=$6, 17, and 53 MD steps, respectively.
The arrow in (f) at $\tau/t_{h} \approx 30$ indicates a local
maxima in P($\tau/t_{h}$) for the two samples with lower
density of pins.  Panel (g) schematically indicates the notation
used in the figures of this work.
}
\label{fig:6}
\end{figure}

In order to directly relate vortex velocities to avalanche 
lifetimes, we consider the distance an individual vortex
moves during an avalanche event.  A vortex normally hops from one 
pinning site to an adjacent site
during the event.  The distance it travels is simply $d_{p}$, 
the average distance between pinning sites,
\begin{eqnarray}
d_{p}=\frac{1}{\sqrt{n_{p}}} \ .
\end{eqnarray}
If we define a ``string'' avalanche as an event during which
each vortex in a chain extending the length of the sample hops
from one site to the next, we can estimate a ``string'' lifetime.
First, we designate the time that a vortex with just enough energy
to depin spends moving between pins as the hopping time,
$t_{h}$.
\begin{eqnarray}
t_{h}=\frac{d_{p}}{v_{c}}\approx\frac{\eta}{f_{p}\sqrt{n_{p}}} \ .
\end{eqnarray}
For example, in PbIn, $\eta \sim 3.3 \times 10^{-8}$ G$^{2}$-s 
\cite{32}, so using
$f_{p} = 3.0 f_{0}$, $n_{p} = 5.93/\lambda^{2}$, and
$\lambda \sim 65$ nm gives $t_{h} \sim 15$ fs.
Next, we estimate $N_{h}$, the number of vortices that must hop in 
order for the avalanche to span the sample:
\begin{eqnarray}
N_{h}=L_{x} \sqrt n_{v} \ ,
\end{eqnarray}
where $L_{x}$ is the sample length.
If we assume that the vortices hop one at a time in sequence,
we obtain an estimate of the ``string'' avalanche lifetime,
\begin{eqnarray}
\tau_{\rm est}=t_{h} N_{h}\approx 
\frac{\eta L_{x}\sqrt{n_{v}}}{f_{p}\sqrt{n_{p}}} \ .
\end{eqnarray}
For instance, using the value of $t_{h}$ for PbIn given above and
taking the vortex density to be $n_{v} = 2.5/\lambda^{2}$ gives a lifetime
of $\tau_{\rm est} \sim 0.6$ ps. 
The relationship between $\tau_{\rm est}$ and $f_{p}$ and $n_{p}$ 
is not simple because the vortex density $n_{v}=n_{v}(H,f_{p},n_{p})$.
For a given field strength, however, it is clear that avalanche
lifetimes decrease as pinning strength or density is increased.  
This is confirmed in the plot of
the distribution $P(\tau)$ in Fig.~\ref{fig:6}(a--b),
where we see that samples with strong dense pinning
have significantly shorter avalanche lifetimes than samples with
weaker pinning [Fig.~\ref{fig:6}(a)]
or samples with lower pinning density
in which interstitial pinning is important 
[Fig.~\ref{fig:6}(b)].
All histograms have been smoothed as described
in Ref. \cite{33}.

In order to determine what fraction of the avalanches
have lifetimes on the order of the estimated ``string'' 
lifetime $\tau_{\rm est}$, we scale the avalanche lifetimes by 
$\tau_{\rm est}$, and plot
$P(\tau/\tau_{\rm est})$ in Fig.~\ref{fig:6}(c--d). 
For the dense strong pinning case of Fig.~\ref{fig:6}(c), 
most avalanche lifetimes are 
shorter than the estimated lifetime $\tau_{\rm est}$ 
($\tau/\tau_{\rm est} < 1$)
as a result of two factors.  First, several vortices hop 
simultaneously, as in Fig.~\ref{fig:4}, 
rather than hopping one after the other, as assumed in our estimate.  
Second, although Fig.~\ref{fig:3} illustrated several system-spanning
avalanches, many events do not involve enough vortices to 
span the entire sample but contain only
a short chain of vortices moving a short distance.
The majority of avalanches 
in samples with strong pinning, $f_{p}^{\rm max}=3.0f_{0}$, contain
less than ten vortices traveling
in short chains that originate in regions of
high vortex density but do not extend into areas with lower vortex density.

\begin{figure}
\centerline{
\epsfxsize=3.5 in
\epsfbox{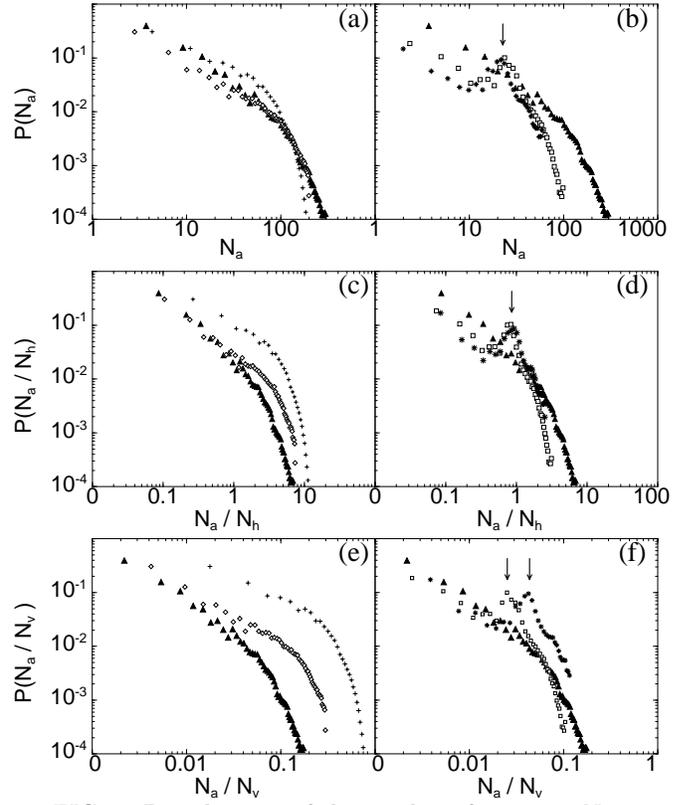}}
\caption{Distributions of the number of vortices, $N_{a}$, moving during
an avalanche event.
The left panels (a,c,e) correspond to 
samples with a high pin density, $n_{p}=5.93/\lambda^{2}$, 
and differing pinning strengths:
filled triangles, $f_{p}^{\rm max}=3.0 f_{0}$;  
open diamonds, $f_{p}^{\rm max}=1.0 f_{0}$;
plus signs, $f_{p}^{\rm max}=0.3 f_{0}$. 
The right panels
(b,d,f) correspond to samples with strong pins, $f_{p}^{\rm max}=3.0 f_{0}$,
and differing pinning densities:
filled triangles, $n_{p}=5.93/\lambda^{2}$;
open squares, $n_{p}=2.40/\lambda^{2}$;
asterisks, $n_{p}=0.96/\lambda^{2}$.  
(c--d) Distributions scaled by the average number of vortices, $N_{h}$,
that must hop for the avalanche to cross the sample.
In (d), note the appearance of a characteristic value in the latter 
two samples at $N_{a} / N_{h} \sim 1$.
(e--f) Distributions scaled by
the average number, $N_{v}$, of vortices in the sample.
For samples with a large density of strong pins,
the leftmost curve in (e) indicates that most avalanches involve
only a tiny fraction of the vortices, resulting in plastic
transport that occurs in {\it brief, choppy bursts}.
}
\label{fig:7}
\end{figure}

As the pinning force is lowered, the chains of moving vortices become
longer on average, leading to an increase in the average value of 
$\tau/\tau_{\rm est}$.
Only samples with a reduced pinning density,
shown in Fig.~\ref{fig:6}(d), produce frequent ``string''
avalanches.  In these samples, nearly all avalanches occur in  
a single easy-flow channel that spans the sample,
as in Fig.~\ref{fig:5}(a), or in two main channels as
in Fig.~\ref{fig:5}(b).
As a result, we observe an increased
likelihood for the ``string'' avalanche in the form of an increase in
the distribution function near $\tau/\tau_{\rm est} \approx 1$.
This increase is most pronounced for the lowest pin density, 
$n_{p}=0.96/\lambda^{2}$ [Fig.~\ref{fig:6}(d)],
where the channel is most strongly established.  
Since the channels in such samples wind significantly in crossing 
the sample, the number of vortices moving during the avalanche is greater
than the estimated value $N_{h}$, and
the observed increase in $P(\tau/\tau_{\rm est})$ 
for $n_{p}=0.96/\lambda^{2}$ falls at
$\tau/\tau_{\rm est} \approx 1.3$, 
rather than at $\tau/\tau_{\rm est} \approx 1$.

Figures~\ref{fig:6}(e--f) show the scaling of the avalanche 
lifetimes by the hopping time $t_{h}$.  In this case, the
distributions $P(\tau/t_{h})$ for 
high pinning density ($n_{p}=5.93/\lambda^{2}$),
in Fig.~\ref{fig:6}(e), collapse and   
can be approximated
for small $\tau$ by the form 
$$P\left(\frac{\tau}{t_{h}}\right) \ \sim \ \left(\frac{\tau}{t_{h}}\right)^{-\gamma} \ ,$$
where $$\gamma \sim 1.4 \ .$$  
Distributions generated by samples with lower pinning densities,
in Fig.~\ref{fig:6}(f),
do not scale and cannot be represented by the same form.
This indicates the importance of the nature of
avalanche propagation.  For
samples with a high density of pins, 
pin-to-pin vortex motion
occurs throughout the sample, with all regions of the sample participating
in an avalanche at some time.
Once the pinning density decreases, however, interstitially pinned
vortices appear
and dominate avalanche transport. 
The vortices always follow the {\it same} paths, visiting as many
interstitial sites as possible when moving through
the sample, and introducing a characteristic avalanche lifetime. 

\subsection{Number of vortices in each avalanche}

Avalanches can be characterized according to the number of vortices
displaced during the event.  This quantity can be directly observed
through computer simulations, but at present must be inferred from
experiments \cite{11}.
We define the number of vortices, $N_{a}$, that were 
actively moving participants in each avalanche
as the number of vortices that were depinned.
Figure~\ref{fig:7}(a--b) presents the 
corresponding distribution $P(N_{a})$.
For high pinning density, as in Fig.~\ref{fig:7}(a),
we can approximate the
distribution for small $N_{a}$ 
as 
$$P(N_{a}) \sim N_{a}^{-\beta} \ . $$
We find that $\beta$ decreases as the pinning strength decreases: 
\[\beta \ \sim \ \left\{ \begin{array}{ll}
1.4 \ , & \  f_{p}^{\rm max}=3.0f_{0} \\
1.0 \ , & \  f_{p}^{\rm max}=1.0f_{0} \\
0.9 \ , & \  f_{p}^{\rm max}=0.3f_{0} \\
\end{array}
\right . \]
Smaller values of $\beta$ indicate a relative increase in the frequency of 
large avalanches compared to small ones.
The trend in $\beta$ might
appear counterintuitive since the smaller flux gradient present in
systems with lower pinning strength $f_{p}$ 
produces less energetic avalanches that might be
expected to have small values of $N_{a}$.
It is, however, the width of the avalanche disturbance rather
than the magnitude of the flux gradient that is important in
determining $N_{a}$.
In samples with {\it strong pinning}, energetic avalanches occur, but
{\it avalanche width is suppressed} since strongly pinned vortices on either
side of the moving chain are not depinned during the short interval
when vortices are moving very rapidly down the steep vortex density 
gradient.
As the pinning weakens, avalanches become wider when weakly pinned 
vortices adjacent to a slowly moving chain depin and
join the motion down the shallow gradient, resulting in
more events with relatively large values of $N_{a}$.

\begin{figure}
\centerline{
\epsfxsize=3.5 in
\epsfbox{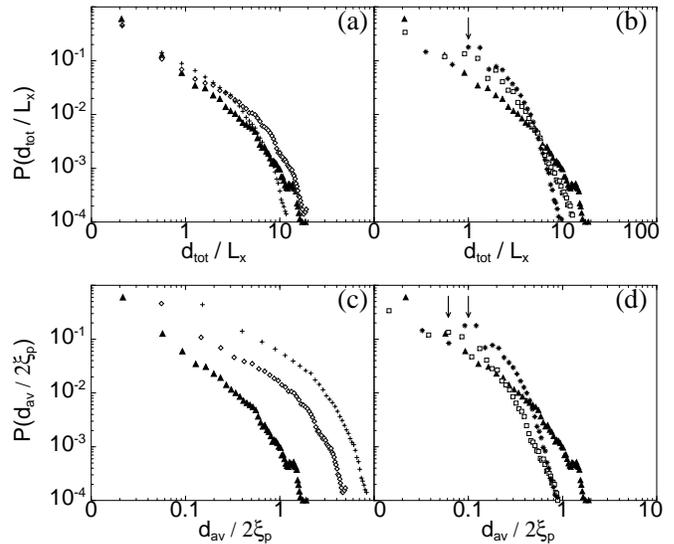}}
\caption{(a--b) Distributions of the total distance, $d_{\rm tot}$, 
moved by all vortices in
the sample scaled by the width of the sample, $L_{x}$.
(c--d) Distributions of the average distance, $d_{\rm av}$,
traveled by a vortex during an avalanche event scaled by
the pin diameter, $2\xi_{p}$.
The left panels (a,c) refer to samples with high pin
density, $n_{p}=5.93/\lambda^{2}$, 
and differing pinning strengths:
filled triangles, $f_{p}^{\rm max}=3.0 f_{0}$;
open diamonds, $f_{p}^{\rm max}=1.0 f_{0}$;
plus signs, $f_{p}^{\rm max}=0.3 f_{0}$. 
The right panels (b,d) refer to samples with 
strong pinning, $f_{p}^{\rm max}=3.0 f_{0}$ ,
and differing pinning densities:
filled triangles, $n_{p}=5.93/\lambda^{2}$;
open squares, $n_{p}=2.40/\lambda^{2}$;
asterisks, $n_{p}=0.96/\lambda^{2}$.  
}
\label{fig:8}
\end{figure}

Vortices flow only in certain channels when $n_{p}$ is low,
leading to the appearance of a characteristic value 
in $P(N_{a})$ [arrow in Fig.~\ref{fig:7}(b)]
and making it impossible to describe
$P(N_{a})$ by a power law form
over the entire range of observed $N_{a}$ values. 
The formation of avalanches with large values of $N_{a}$ is
impeded in such cases because all available interstitially pinned
vortices are already participating in the channel, and the
neighboring strongly pinned vortices cannot be depinned.

The arrow in Fig.~\ref{fig:7}(b) indicates a value of $N_{a}$
corresponding
to the number of vortices in the channels $N_{\rm channel}$.
Since the channels span the sample, we expect 
$N_{\rm channel}$ to be close to the number of vortices $N_{h}$ that
must hop to cross the sample in a straight chain, 
where $N_{h}=L_{x} \sqrt n_{v}$.
To verify this, we plot the scaled quantity 
$P(N_{a}/N_{h})$ in Fig.~\ref{fig:7}(c--d).
We find in Fig.~\ref{fig:7}(d) that samples with low pin 
density $n_{p}$, in
which interstitial pinning is important, have
$N_{\rm channel} \sim N_{h}$,
as is expected if all moving vortices form a single chain across
the sample.  Specifically, increases in $P(N_{a}/N_{h})$ fall at
$N_{a}/N_{h} \approx 1$ for 
$n_{p}=2.40/\lambda^{2}$ and $N_{a}/N_{h} \approx 1.1$ for 
$n_{p}=0.96/\lambda^{2}$.
The winding of the channels in samples with $n_{p}=0.96\lambda^{2}$
causes the value of $N_{a}$ corresponding to 
channel motion to be slightly larger than $N_{h}$.
For samples with high pinning density,
Fig.~\ref{fig:7}(c) indicates that the most linear part of
$P(N_{a}/N_{h})$ is produced for 
$N_{a} \lesssim N_{h}$, or for events
that did not cross the entire sample.  
Broad avalanches are more than one vortex chain wide and have
$N_{a}/N_{h} > 3$.  These avalanches 
are more common in samples with smaller $f_{p}$, as indicated 
by the rightmost curve
in Fig.~\ref{fig:7}(c), where
samples with weak pinning provide the largest avalanches
for a fixed probability density $P(N_{a}/N_{h})$.

To determine how effectively avalanches transport vortices
across the sample, we scale $N_{a}$ by the
total number of vortices $N_{v}$ present in the sample, and plot the result
in Fig.~\ref{fig:7}(e--f).
Only a tiny fraction of the
vortices participate in a typical event
when pinning is strong and dense, as indicated by 
the leftmost curve in Fig.~\ref{fig:7}(e).
The avalanches in this sample are also very short lived, as already shown in
Fig.~\ref{fig:6}(a),
so this sample is best characterized by {\it plastic transport}
that occurs in {\it brief, choppy bursts}.
On the other hand, for
a high density of weak pinning sites, 
we find that it is possible for a significant portion
of the vortices in the sample to collectively move in an avalanche, as in
the rightmost curve of Fig.~\ref{fig:7}(e). 
Vortex motion in these long-lived avalanches
is less plastic, with 
small adjacent portions of the vortex lattice gradually sliding forward 
at different times. 
The samples with low pin density, shown in Fig.~\ref{fig:7}(f),
have their transport dominated by the single easy-flow channel that
develops.  We find that the fraction of vortices contained in this
channel increases as pinning density decreases since the interstitial
channel winds to a greater degree in the sample with fewer pins
and involves a correspondingly larger number of vortices.

\subsection{Total vortex displacement}

Another measure of avalanche size, uniquely available
through simulation, is the total displacement $d_{\rm tot}$
of all vortices in the sample during the event,
\begin{eqnarray}
d_{\rm tot}=\sum_{i=1}^{N_{v}} d_{i} \ ,
\end{eqnarray}
where $d_{i}$ is the displacement of each vortex.
In order to find the probability of an avalanche that spans the sample length
with $d_{\rm tot} \gtrsim L_{x}$, we plot $P(d_{\rm tot}/L_{x})$ in
Fig.~\ref{fig:8}(a--b).
Occasionally we observe avalanches with
the surprisingly large $d_{\rm tot}/L_{x} \sim 10$.
These large values reflect the cumulative effect of vortices
that experience very small displacements inside the parabolic pinning 
sites rather than moving
to a new pinning site.  For instance,
if 800 vortices each move a distance 
$d_{i} \sim \xi_{p}/10=0.015 \lambda$, 
the total displacement recorded would be $12 \lambda$.  Such 
very small vortex displacements
play a very important role in {\it transmitting stress} throughout the 
vortex lattice.
Figure~\ref{fig:8}(a) indicates that $P(d_{\rm tot}/L_{x})$ is
very robust against variations in pinning strength.  Excluding
large values of $d_{\rm tot}/L_{x}$, the distributions can
be approximated by the form 
$$P\left(\frac{d_{\rm tot}}{L_{x}}\right) \ \sim \ \left(\frac{d_{\rm tot}}{L_{x}}\right)^{-\delta} \ .$$
We find that $\delta$ decreases as the pinning strength decreases: 
\[\delta \ \sim \ \left\{ \begin{array}{ll}
1.7 \ , & \  f_{p}^{\rm max}=3.0f_{0} \\
1.2 \ , & \  f_{p}^{\rm max}=1.0f_{0} \\
1.1 \ , & \  f_{p}^{\rm max}=0.3f_{0} \\
\end{array}
\right . \]
Figure~\ref{fig:8}(b) indicates that $P(d_{\rm tot}/L_{x})$ is
weakly dependent on $n_{p}$. 
The effect of the channels appears as an enhancement of the distribution
near $d_{\rm tot}/L_{x}=1$, marked by an arrow in Fig.~\ref{fig:8}(b).

We can compare the relative amount of vortex transport
in different samples by considering the
average distance $d_{\rm av}$ moved by a vortex during an avalanche,
\begin{eqnarray}
d_{\rm av}=\frac{1}{N_{v}} \sum_{i=1}^{N_{v}} d_{i}= \frac{d_{\rm tot}}{N_{v}} \ .
\end{eqnarray}
Since most vortices do not become depinned, we plot 
$P(d_{\rm av}/2\xi_{p})$, where $2\xi_{p}$ is the diameter of 
the pinning trap.
The plot of $P(d_{\rm av}/2\xi_{p})$ in
Fig.~\ref{fig:8}(c) shows that as the pinning strength $f_{p}$ is
reduced, $d_{\rm av}/2\xi_{p}$ increases, indicating that 
individual vortices are likely to travel further during
an avalanche.  This occurs both because 
a large fraction of vortices in samples with weak pinning
participate in the avalanche, as demonstrated in Fig.~\ref{fig:7}(e),
and because the average spacing between vortices increases since vortex
density decreases with decreasing pinning strength.
In Fig.~\ref{fig:8}(d), we find that reducing the
pinning density $n_{p}$ can either increase or decrease $d_{\rm av}$.
As $n_{p}$ decreases, the strength of the interstitial pinning sites
decreases, allowing individual vortices to travel further. 
Not as many vortices are moving, however,
since the motion is restricted to well defined channels.
These two competing effects enhance $P(d_{\rm av})$ at small values of
$d_{\rm av}$, but reduce it at large values.

\begin{figure}
\centerline{
\epsfxsize=3.5 in
\epsfbox{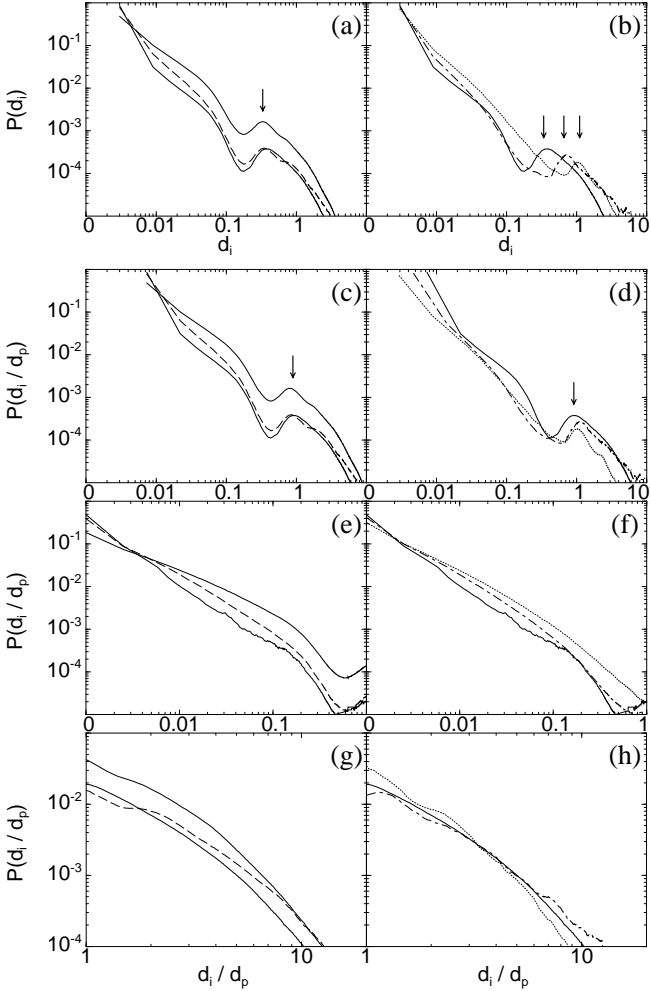}}
\caption{(a--b) Distributions of the distance, $d_{i}$, 
traveled by individual vortices during all the  avalanche events.
Characteristic displacements $d_{h}$ are marked with arrows.
(c--d) Distributions scaled by
the inter-pin distance, $d_{p}$.
Only vortices that moved more than $1.0 d_{p}$ were directly participating
in the avalanche.  The rest were shifting
very slightly in pinning wells and transmitting
stress through the lattice.
(e--f) Distributions $P(d_{i}/d_{p})$ for only those vortices 
that traveled a distance $d_{i}<d_{p}$, i.e., vortices that were
not depinned.
(g--h) Distributions $P(d_{i}/d_{p})$ for only those vortices that traveled
a distance $d_{i}>d_{p}$, i.e., vortices that were depinned.
The left panels (a,c,e,g) correspond to 
samples with a high density of pins, $n_{p}=5.93/\lambda^{2}$, 
and differing pinning strengths:
solid line, $f_{p}^{\rm max}=3.0 f_{0}$;
dashed line, $f_{p}^{\rm max}=1.0 f_{0}$;
heavy solid line, $f_{p}^{\rm max}=0.3 f_{0}$. 
The right panels
(b,d,f,h) correspond to samples with strong pinning,
$f_{p}^{\rm max}=3.0 f_{0}$, 
and differing pinning densities:
solid line, $n_{p}=5.93/\lambda^{2}$;
dot-dashed line, $n_{p}=2.40/\lambda^{2}$;
dotted line, $n_{p}=0.96/\lambda^{2}$.  
}
\label{fig:9}
\end{figure}

\subsection{Individual vortex displacements}

The microscopic information available in our simulation permits
us to calculate individual vortex displacements.  
In Fig.~\ref{fig:9} we plot 
$P(d_{i})$, where $d_{i}$ is the distance a single vortex
is displaced in an avalanche.  In every distribution,
we find that
vortex motion can be described in terms of a characteristic
size $d_{h}$, indicated
by arrows in the figure.  This size roughly corresponds to the average distance
between pinning sites $d_{p}$,
\begin{eqnarray}
d_{h}\approx d_{p} =\frac{1}{\sqrt n_{p}} \ ,
\end{eqnarray}
and appears because vortices participating in an
avalanche typically hop from one pinning site to an adjacent
pinning site.
Thus, in Fig.~\ref{fig:9}(a), $d_{h}$ falls at the
same value for three samples with equal pinning density $n_{p}$,
while in Fig.~\ref{fig:9}(b), $d_{h}$ increases
as $n_{p}$ decreases.
The vortices in the sample can thus be divided into three
categories:  those that did not leave a pinning site, 
$d_{i}<2\xi_{p}<d_{h}$, those that hopped from
one pinning site to another, $d_{i} \sim d_{h}$,
and those that hopped more than one pinning site,
$d_{i} > d_{h}$.

We first consider those vortices that hopped one pinning site.
If we scale the distributions by the inter-pin distance
$d_{p}$, as in 
Fig.~\ref{fig:9}(c--d),
we find that the characteristic distances 
in $P(d_{i}/d_{p})$ fall
at $d_{i}/d_{p} \sim 1$ (marked with an arrow), with
a second smaller increase at $d_{i}/d_{p} \sim 2$ occurring
in samples where vortices occasionally hopped two
pinning sites.

Next we examine the motion of vortices shifting very
slightly inside pinning
wells.  
In Fig.~\ref{fig:9}(e--f), we
focus on vortices that
did not leave a pinning site, $d_{i}<d_{p}$, and present 
$P(d_{i}/d_{p})$.  We find that, for small values of
$d_{i}/d_{p}$, these distributions
can be described by the form 
$$P\left(\frac{d_{i}}{d_{p}}\right) \sim \left(\frac{d_{i}}{d_{p}}\right)^{-\rho} \ ,$$ 
where:
$\rho \approx 1.4$ for $n_{p}=5.93/\lambda^{2}$ and 
$f_{p}^{\rm max}=3.0f_{0}$;
$\rho \approx 1.4$ for $n_{p}=5.93/\lambda^{2}$ and
$f_{p}^{\rm max}=1.0f_{0}$;
$\rho \approx 1.4$ for $n_{p}=2.40/\lambda^{2}$ and
$f_{p}^{\rm max}=3.0f_{0}$;
$\rho \approx 1.2$ for $n_{p}=0.96/\lambda^{2}$ and
$f_{p}^{\rm max}=3.0f_{0}$; and
$\rho \approx 0.9$ for $n_{p}=5.93/\lambda^{2}$ and
$f_{p}^{\rm max}=0.3f_{0}$.
The two samples with smaller values of $\rho$ also
have the lowest critical
currents $J_{c}$, as determined from the slope of the flux profiles
in Fig.~\ref{fig:1}.

A simple two-dimensional argument shows that
$\rho$ is expected to be similar (and close to one) for all
samples, since only vortices that experience extremely small displacements
in pinning sites
affect the value of $\rho$.  
If we add an additional vortex to an already existing
arrangement of vortices,  the repulsive force, $f_{\rm extra}$,
from this vortex
will cause each nearby vortex to move a very small distance $u_{i}$ in its
pinning well in order to reach a new equilibrium position.  
Neglecting higher order effects, we find that
the very small displacement $u_{i}$ of the $i$th vortex 
is determined by the distance from the added
vortex and by the form of the force $f_{\rm extra}$.
We take the distance from the perturbing vortex to be $r$,
and assume that the additional force is applied during a time $\delta t$.
The vortex-vortex interaction force gives
\begin{eqnarray}
f_{\rm extra}=\frac{\Phi_{0}^{2}}{8\pi^{2}\lambda^{3}} \, K_{1}\!\left(\frac{r}{\lambda}\right)
\end{eqnarray}
If we consider the limit of small $r$, $r \ll \lambda$, we can use
the asymptotic form of the Bessel function to write
\begin{eqnarray}
f_{\rm extra} \approx \frac{\Phi_{0}^{2}}{8\pi^{2}\lambda^{3}} \frac{\lambda}{r} \ .
\end{eqnarray}
We have (for $\delta u < \xi_{p} \ll \lambda$)
\begin{eqnarray}
\delta u(r) \; = \; \frac{f_{\rm extra}}{\eta} \delta t \; \approx \;
\frac{\delta t}{\eta} \frac{\Phi_{0}^{2}}{8\pi^{2}\lambda^{3}} \frac{\lambda}{r} \ .
\end{eqnarray}
To obtain a distribution, 
we find the number of vortices $N(r)$ located a
radius $r$ from the perturbing vortex.  Due to the cylindrical symmetry,
this can be written simply as
\begin{eqnarray}
\delta N(r)=2\pi r n_{v} \ \delta r \ ,
\end{eqnarray}
where $\delta r$ is an infinitesimal displacement.
We want an expression for the slope of our distributions,
\begin{eqnarray}
\rho=-\frac{d}{d(\ln(\delta u))} \ln \delta N \ .
\end{eqnarray}
Thus, we write:
\begin{eqnarray}
\ln \delta N(r)=\ln(2 \pi n_{v} \delta r) + \ln r
\end{eqnarray}
\begin{eqnarray}
\frac{d \ln \delta N(r)}{dr}=\frac{1}{r}
\end{eqnarray}
\begin{eqnarray}
\ln \delta u(r) \approx \ln \left ( \frac{\Phi_{0}^{2}\delta t}{8\pi^{2}\lambda^{2}\eta} \right ) - \ln r
\end{eqnarray}
\begin{eqnarray}
\frac{d \ln \delta u}{dr} \approx -\frac{1}{r} \ .
\end{eqnarray}
Therefore, we find 
\begin{eqnarray}
-\rho = \frac{d \ln \delta N}{d \ln \delta u} \approx \left(\frac{1}{r}\right) \left(-\frac{1}{r}\right)^{-1}=-1 \ .
\end{eqnarray}
For small displacements
($\delta u < \xi_{p} \ll \lambda$), this predicts $\rho \sim 1$, in general
agreement with our observed values ($\rho \approx 0.9 - 1.4$).

The argument presented above is independent of the
shape of the pinning potential, and assumes that the vortex density $n_{v}$ is
constant throughout the sample.  The presence of the Bean critical state
makes this assumption inaccurate.  It can, however, be shown that including
the critical state does not significantly
affect the expected value for the slope.
We describe the field in the sample according to the Bean model,
\begin{eqnarray}
B=H_{\rm ext} - \frac{2\pi J_{c}r}{c} \ ,
\end{eqnarray}
where $H_{\rm ext}$ is the externally applied field.  The field can
also be written as
\begin{eqnarray}
B=n_{v}\Phi_{0} \ .
\end{eqnarray}
Thus,
\begin{eqnarray}
n_{v}(r)=\frac{H_{\rm ext} - 2\pi J_{c}r/c}{\Phi_{0}} \ .
\end{eqnarray}
The number of vortices located at a distance $r$ becomes
\begin{eqnarray}
\delta N(r)=2\pi r \frac{(H_{\rm ext} - 2\pi J_{c}r/c)}{\Phi_{0}} \delta r \ .
\end{eqnarray}
We then have
\begin{eqnarray}
\ln \delta N(r)=\ln \left(\frac{2\pi \delta r}{\Phi_{0}}\right) + \ln r + \ln (H_{\rm ext} - 2\pi J_{c}r/c)
\end{eqnarray}
\begin{eqnarray}
\frac{d \ln \delta N(r)}{dr}=\frac{1}{r} + \frac{-2\pi J_{c}/c}{H_{\rm ext} - 2\pi J_{c}r/c} \ .
\end{eqnarray}
The external field $H_{\rm ext}$ 
can be written in terms of the critical current $J_{c}$
and the distance the field has penetrated into the sample, $r^{*}$:
\begin{eqnarray}
H_{\rm ext}=\frac{2\pi J_{c} r^{*}}{c}
\end{eqnarray}
We obtain
\begin{eqnarray}
\frac{d \ln \delta N}{d \ln \delta u}  \approx -1 - \frac{-r}{r^{*} - r}= -1 - \frac{1}{1-(r^{*}/r)} \ .
\end{eqnarray}
We have $r^{*} \gg r$, since the field has penetrated the entire sample
length $L_{x} \gg \lambda$.  We therefore write
\begin{eqnarray}
\frac{ d \ln \delta N }{ d \ln \delta u } \approx -1 - \left(\frac{r}{r^{*}}\right ) \frac{1}{1 - (r/r^{*})} \approx -1 - \frac{r}{r^{*}} \ .
\end{eqnarray}
Thus, including the Bean gradient introduces only a negligible
correcting term $r/r^{*}$ to the expected slope, since $r/r^{*}\ll 1$.

It is interesting to consider what slope value would be obtained if
the vortex-vortex interaction were that found in a thin film, rather
than the form for a thick slab of material used above.  In this case,
the additional force from a vortex is
\begin{eqnarray}
f_{\rm extra} \sim \frac{1}{r^{2}}
\end{eqnarray}
giving
\begin{eqnarray}
\delta u(r) \sim \frac{1}{\eta r^{2}} \delta t
\end{eqnarray}
\begin{eqnarray}
\ln \delta u(r) \sim \ln \frac{\delta t}{\eta} -2 \ln r
\end{eqnarray}
\begin{eqnarray}
\frac{d \ln \delta u}{dr} \sim -\frac{2}{r}
\end{eqnarray}
\begin{eqnarray}
\frac{d \ln \delta N}{d \ln \delta u} \sim \left(\frac{1}{r}\right) \left(-\frac{2}{r}\right)^{-1}=-\frac{1}{2} \ .
\end{eqnarray}
Thus, the distribution of individual displacements in a thin film 
should differ from that produced by a slab of material
($\rho \sim 0.5$ versus $\rho \sim 1$).

In Fig.~\ref{fig:9}(g--h), we focus on the displacements of 
vortices that were actively participating in avalanches, and plot
$P(d_{i}/d_{p})$ for vortices with $d_{i} > d_{p}$.
We find that decreasing the pinning
strength leads to a very slight  overall increase in the distance traveled
by a moving vortex, as indicated by the slightly higher likelihoods for moving
larger distances in Fig.~\ref{fig:9}(g).  This is a result of
the fact that, in general, avalanche disturbances are larger and longer
lived in these samples, allowing individual vortices the opportunity
to travel two or more pinning sites in an avalanche.
Decreasing the pinning density, as in Fig.~\ref{fig:9}(h),
also affects the likelihood that a vortex will move a certain distance.  

\begin{figure}
\centerline{
\epsfxsize=3.5 in
\epsfbox{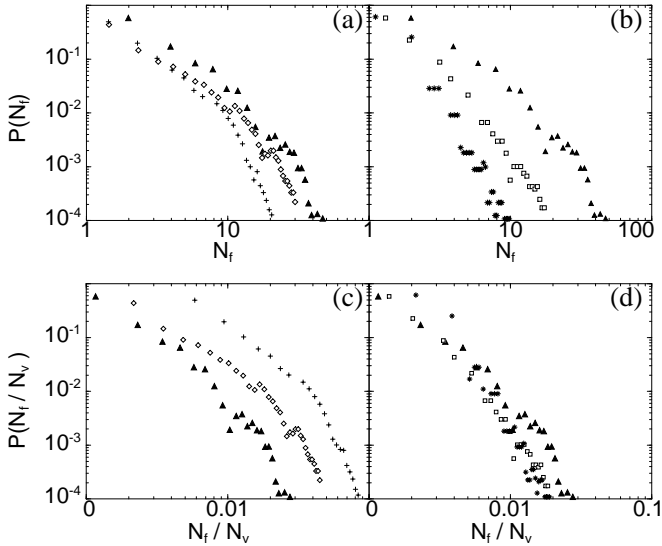}}
\caption{(a--b) Distributions of the number $N_{f}$ 
of vortices falling off the edge of the sample.
(c--d) Distributions scaled by the average total number $N_{v}$
of vortices in the sample.
The left panels (a,c) correspond to  samples with 
a high density of pins, $n_{p}=5.93/\lambda^{2}$, 
and differing pinning strengths:
filled triangles, $f_{p}^{\rm max}=3.0 f_{0}$;
open diamonds, $f_{p}^{\rm max}=1.0 f_{0}$;
plus signs, $f_{p}^{\rm max}=0.3 f_{0}$.  
When $N_{f}$ is not very large, 
the slope of the curve in each case is
roughly $-2.4$.
The right panels
(b,d) correspond to samples with strong pinning, $f_{p}^{\rm max}=3.0 f_{0}$, 
and differing pinning densities:
triangles, $n_{p}=5.93/\lambda^{2}$;
squares, $n_{p}=2.40/\lambda^{2}$;
asterisks, $n_{p}=0.96/\lambda^{2}$.  
For small values of $N_{f}$, the slope of the curves in (b) is
roughly $-2.4$, $-3.4$, and $-4.4$, respectively.
}
\label{fig:10}
\end{figure}

\subsection{Vortices leaving the sample}

The number of vortices $N_{f}$ that exit the sample during an event
can be directly compared with values obtained in experiments, both in vortex
systems \cite{11,12,13} and in sandpiles
\cite{1,2,3}.
In Fig.~\ref{fig:10}(a--b) we plot $P(N_{f})$.
Events with $N_{f}=0$ are not shown.  The relatively
small size of our sample resulted in a smaller data set for this
quantity than for all other quantities considered in this paper.
If we approximate
the distributions for small $N_{f}$ by the form 
$$P(N_{f}) \sim N_{f}^{-\alpha} \ ,$$
we find that all samples with high pinning density have 
$$\alpha \sim 2.4 \ .$$  
As the pinning density $n_{p}$ is decreased for constant $f_{p}$,
$\alpha$ increases, with $\alpha \sim 3.4$ for $n_{p}=2.40/\lambda^{2}$
and $\alpha \sim 4.4$ for $n_{p}=0.96/\lambda^{2}$.
A larger $\alpha$ indicates a relatively smaller likelihood for
the occurrence of events with large $N_{f}$.

When we examine the region from which vortices exit the sample,
we find that flux can exit from any location in samples with high pin density,
as seen from the relatively uniform coverage of trails along the right (outer)
edge of Fig.~\ref{fig:5}(c) and Fig.~\ref{fig:5}(d).  
Even after one location has been depleted by a large avalanche, other
areas along the sample edge still contain enough vortices to
remain active in large and small events 
while the depleted regions refill.
As the pinning density is decreased
and interstitial channels develop, avalanche paths
become highly constrained and flux exits in only a few locations
that have difficulty building up enough stress to permit
events with large $N_{f}$ to occur.
For example, with the reduced number of flux paths in Fig.~\ref{fig:5}(b), 
events with large $N_{f}$ are rare, and thus
$\alpha$ is large: $\alpha \sim 3.4$.
Fig.~\ref{fig:5}(a) illustrates a small region of 
the extreme case of a sample with only one exit channel, for
which $n_{p}=0.96/\lambda^{2}$ and $\alpha$ has the larger value 
$\alpha \sim 4.4$.  Large values of $N_{f}$ occur only when
this single channel is filled by vortices under high stress.
Since all smaller events happen through this {\it same}
channel, however, stress in the channel is relieved by small events before
it can build to a high level.
Thus, the likelihood that the channel will produce a very large
$N_{f}$ is extremely small.  As expected, the cut-off of
$P(N_{f})$ for this sample falls at a much lower $N_{f}$ than
that of the other samples, as seen in the leftmost curve in 
Fig.~\ref{fig:10}(b).

Strong pinning effectively confines the avalanche disturbance
to a narrow channel
and prevents movement from spreading throughout the sample.  Weak
pinning permits collective motion in larger regions of the sample.
These effects are more pronounced for samples with a lower density
of pinning sites.
This is highlighted by a comparison of how effectively individual avalanches
remove vortices from different samples.  In Fig.~\ref{fig:10}(c--d),
we scale $N_{f}$   by the number of vortices in the sample $N_{v}$.
Most vortices in a sample with strong pinning cannot exit;  those
that do exit move in isolated chains.  
As the pinning strength is weakened, however,
a greater percentage of the vortices are able to exit in events
that consist of the collective motion of several adjacent
chains.  This is illustrated by the rightmost curves in 
Fig.~\ref{fig:10}(c).

\subsection{Individual vortex velocities}

\begin{figure}
\centerline{
\epsfxsize=3.5 in
\epsfbox{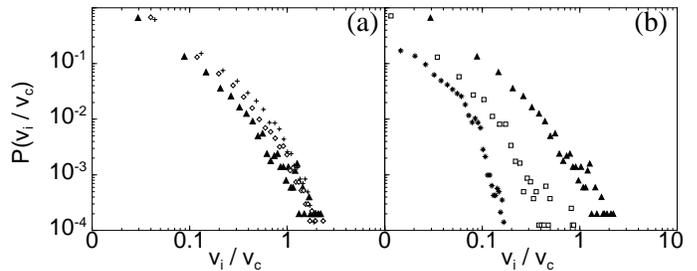}}
\caption{Distributions of the individual velocities, $v_{i}$, of vortices in a
thin strip of the sample,
scaled by the characteristic velocity $v_{c}$.
The left panel (a) corresponds to samples with 
a high pinning density, $n_{p}=5.93/\lambda^{2}$, 
and differing pinning strengths:
filled triangles, $f_{p}^{\rm max}=3.0 f_{0}$;
open diamonds, $f_{p}^{\rm max}=1.0 f_{0}$;
plus signs, $f_{p}^{\rm max}=0.3 f_{0}$.
The right panel (b) corresponds to samples with 
strong pinning, $f_{p}^{\rm max}=3.0 f_{0}$, 
and differing pinning densities:
triangles, $n_{p}=5.93/\lambda^{2}$;
squares, $n_{p}=2.40/\lambda^{2}$;
asterisks, $n_{p}=0.96/\lambda^{2}$.
}
\label{fig:11}
\end{figure}

Due to the presence of a non-uniform magnetic flux
gradient that is not of the ideal Bean form,
the velocity of an average vortex is a function of position
(see Fig.~\ref{fig:1}).
Therefore, to construct a distribution of individual
vortex velocities $v_{i}$ that is not affected by the gradient, 
we select a narrow region
of the sample where the gradient is essentially constant and
find $P(v_{i})$ for vortices in this region.  
The individual velocities $v_{i}$ are then scaled
by the characteristic velocity $v_{c}$, given by
\begin{eqnarray}
v_{c}=\frac{d_{p}}{t_{h}} \approx \frac{f_{p}}{\eta} \ ,
\end{eqnarray}
where $d_{p}$ is the average distance between pinning sites, and $t_{h}$ is
the average hopping time,
$t_{h}=\eta d_{p}/f_{p}.$  Figure~\ref{fig:11} 
presents the resulting distributions $P(v_{i}/v_{c})$.
For a high density of pinning [Fig.~\ref{fig:11}(a)], 
we find that pinning strength changes $v_{c}$ but does not affect the 
form of $P(v_{i}/v_{c})$.
Note that vortices moving at speeds less
than $v_{c}$ do not have sufficient energy to escape from a pinning site
and are merely experiencing very small displacements inside
their respective pinning wells.  The similar form of
$P(v_{i}/v_{c})$ is then explained by the fact that 
all of the wells are parabolic.
For the lowest velocities, we can approximate $P(v_{i}/v_{c})$
with
$$P\left(\frac{v_{i}}{v_{c}}\right) \sim \left(\frac{v_{i}}{v_{c}}\right)^{-\sigma} \ ,$$
where $$\sigma \sim 2 \ .$$
Different densities of pinning wells alter the form of $P(v_{i}/v_{c})$,
as can be seen from Fig.~\ref{fig:11}(b).  As the pinning
density decreases, some vortices experience small displacements
in interstitial wells while others remain in parabolic
traps.  The combined effect changes the overall distribution.

If, instead of considering the distribution of individual 
velocities, we construct the power spectrum of the velocity 
signal $v_{\rm av}$, we find that all of the samples with a 
high pinning density produce broad power spectra $S(f)$ of 
the form $S(f) \sim f^{-\nu}$, where
$1.5 < \nu < 2.0$. Samples with lower pinning density do 
not produce spectra of this form.  The noise power spectra 
will be discussed further elsewhere \cite{34}.

\section{Recent experiments on vortex plastic flow and avalanches}

In several experiments, the nature of the vortex movement has been inferred 
indirectly by using, for instance, 
voltage signals, resulting in several plausible scenarios
of vortex motion including cylindrical
bundles \cite{17}, clump-like bundles \cite{32,35},
or elongated bundles \cite{36}.
Due to the presence of a field gradient in the sample,
we find that
{\it all collectively moving groups of vortices are arranged in
one or more long, narrow moving chains}
rather than clumped or cylindrical bundles.
The existence of quasi-one-dimensional vortex channels is suggested
in Refs. \cite{32,37}, where
vortex motion is assumed to occur in {\it straight} paths 
that cross the sample.
The chain-like channels that we observe resemble these suggestions,
except that in our simulation the vortex paths often {\it wind}
through the sample rather than following the shortest path
down the gradient.  In addition, our simulation indicates that the
velocity of the moving vortices during a single avalanche is 
not constant but is pulse-like,
that is, with oscillations in the average velocity $v_{\rm av}$ during the
vortex avalanche.

In the experiment by Field {\it et al.} \cite{11}, 
flux lines exiting a superconducting tube were detected with a coil.
Our average velocity signals, plotted in Fig.~\ref{fig:2}, strongly
resemble the resulting voltage signals in Fig. 1 of Ref. \cite{11}.  
Both simulation and experiment find distributions of avalanche
sizes $N_{f}$ that can be approximated by power laws with a range
of exponents.  The experiment in Ref. \cite{11} finds 
$\alpha \sim 1.4$ to $2.2$, while
we find $\alpha \sim 2.4$ to $4.4$ using a variety of different samples.
The behavior of the experimental sample therefore 
resembles our simulations with
a very high density of pinning sites, $n_{p}=5.93/\lambda^{2}$,
and no characteristic channels.  
This is reasonable since
the density of pinning in the experimental sample is
very high.  The pinning centers are grain
boundaries with spacings randomly distributed between
about 30 and 50 nm, while 
\twocolumn[\hsize\textwidth\columnwidth\hsize\csname
@twocolumnfalse\endcsname
\begin{table}
\begin{tabular} {c|ccccc}\hline
{Quantity} &
{$f_{p}^{\rm max} = 0.3f_{0}$} & {$f_{p}^{\rm max} = 1.0f_{0}$}
& {$f_{p}^{\rm max} = 3.0f_{0}$} & {Ref. [11]} & {Ref. [13]} \\ \hline
{$\tau/t_{h}$} & $\gamma \sim 1.4$ & $\gamma \sim 1.4$ & $\gamma \sim 1.4$ 
& -- & -- \\
{$N_{a}$} & $\beta \sim 0.9$ & $\beta \sim 1.0$ & $\beta \sim 1.4$ 
& -- & -- \\
{$d_{\rm tot}/L_{x}$} & $\delta \sim 1.1$ & $\delta \sim 1.2$
& $\delta \sim 1.7$ & -- & -- \\
{$d_{i}/d_{p}$} & $\rho \sim 0.9$ & $\rho \sim 1.2$ & $\rho \sim 1.4$ 
& -- & -- \\
{$v_{i}/v_{c}$} & $\sigma \sim 2.0$ & $\sigma \sim 2.0$ & $\sigma \sim 2.0$
& -- & -- \\
{$N_{f}$} & $\alpha \sim 2.4$ & $\alpha \sim 2.4$ & $\alpha \sim 2.4$ & 
$\alpha \sim 1.4$ to $2.2$ & $\alpha \sim 1.7$ to $2.2$ \\ \hline
\end{tabular}
\caption{Best-fit slopes of the most linear region of distributions of
several quantities.  The form of the distributions in each
case were:  $P(\tau/t_{h}) \sim (\tau/t_{h})^{-\gamma}$,
$\ P(N_{a}) \sim N_{a}^{-\beta}$, 
$\ P(d_{\rm tot}/L_{x}) \sim (d_{\rm tot}/L_{x})^{-\delta}$, 
$\ P(d_{i}/d_{p}) \sim (d_{i}/d_{p})^{-\rho}$,
$\ P(v_{i}/v_{c}) \sim (v_{i}/v_{c})^{-\sigma}$,
$\ P(N_{f}) \sim N_{f}^{-\alpha}$.  
Notice that a multiplicative factor
in the independent variable does not alter the slope.  Thus, $\tau$ and
$\tau/t_{h}$ produce distributions with 
the same slope.   The samples listed here had a high density
$n_{p} = 5.93/\lambda^{2}$ of pinning sites.
Samples with $f_{p}^{\rm max} = 3.0 f_{0}$ and
lower densities $n_{p}$ of pins did not produce power law
distributions.}
\end{table}
\vskip2pc]

\hspace{-13pt}
$\lambda$ is on the order of
400 nm, leading to a pinning density of $n_{p} \sim 100/\lambda^{2}$.
Thus, this sample has a higher pin density  than our most
densely pinned case, and it is
therefore reasonable to expect the experimental exponents to be less than 2.4
based on the general trend of our results.  

Zieve {\it et al.} \cite{12} observed avalanches at
higher field strengths, where the pinning forces are expected
to be reduced, and interpret their results to mean that the high-field
regime produces large-scale, plastic rearrangements of the vortices,
whereas the low-field regime generates elastic rearrangements.
It is possible that there were undetectable steps in the hysteresis
loops they obtained in the low-field regime since
the smallest flux jumps resolved by their
apparatus were on the order of 0.1 kOe.
If their sample contained either characteristic channels for flux flow
or high strength pinning, only a small fraction of the flux inside the
sample would be transported through narrow exit regions at these
low fields.
As the field density increased, their sample would cross
into the regime 
of weak dense pinning.  In this case, as shown by our
simulations,
the sample can effectively transport much 
more flux, producing larger flux steps that can be resolved
by the experiment.  

Finally, we consider very recent work by Nowak {\it et al.} \cite{13},
who obtain different types of avalanche distributions as a function
of temperature.
At higher temperatures, they find a broad distribution
for the amount of flux exiting the sample in each event, 
$$P(N_{f}) \sim N_{f}^{-\alpha} \ , $$
with $\alpha$ ranging from 1.7 to 2.2, 
in agreement with both Ref. \cite{11}
and our results.
At the lowest temperatures, characteristic
avalanche sizes appear in the form of events involving a large
number of vortices that may be system-spanning.
These changes have been discussed in terms of thermal
instabilities in the material \cite{13}, but
they can also be considered from the standpoint of
the presence or absence of channels for vortex motion.
As we have seen in our simulations, channels are most likely
to form when the vortices are able to move interstitially.
This is expected to happen whenever the pinning force
exceeds the minimum pinning strength $f^{(p)}_{\rm min}$ that
would permit interstitial motion;  
from Eq.~\ref{eqn:interstitial}, this occurs when
\begin{eqnarray}
f_{p} > f^{(p)}_{\rm min} = f_{0} \ K_{1}\! \left( \frac{1}{2\lambda (T)\sqrt{n_{p}}}\right) \ .
\end{eqnarray}
Writing $t = T/T_{c}$,
\begin{eqnarray}
\frac{\lambda (T)}{\lambda (0)}=\frac{1}{\sqrt{1-t^4}} ,
\end{eqnarray}
and the strength of the
pinning force $f_{p}$ required for interstitial motion to occur
increases with temperature:
\begin{eqnarray}
f_{p}>f_{0}\ K_{1}\!\left(\frac{\sqrt{1-t^4}}{2\lambda (0)\sqrt{n_{p}}}\right)\ .
\end{eqnarray}
At the lowest temperatures, $f_{\rm req}$ is small, and
interstitial motion is possible in the sample.
Channels of flux flow form, and a characteristic avalanche
size appears in the 
distribution of avalanche sizes.
As the temperature increases, $f_{\rm req}$ increases until some
of the pinning sites in the sample are no longer strong enough to
permit interstitial flow.  In this case, pin-to-pin vortex motion
will occur evenly throughout the sample, and the distribution of
avalanche sizes will broaden.
We therefore expect a transition from pin-to-pin motion
at higher temperatures to interstitial channel flow at low temperatures,
with a corresponding transition from
broad distributions of $N_{f}$ (for high $T$)
to a characteristic value of $N_{f}$ (for low $T$).
The transition with temperature in the nature of the distributions is
experimentally observed in Ref. \cite{13}.  

Table I presents a summary of our results for the exponents of the power
laws found in our avalanche distributions.  Only the distribution
of flux lines falling off the edge of the sample was measured
experimentally \cite{11,13}.  The other quantities are more
difficult to measure experimentally.  However, novel flux imaging
techniques (such as magneto-optical imaging, arrays of Hall
probes \cite{13,38}, scanning Hall probes, and especially,
Lorentz microscopy) could make it possible to obtain some of these
distributions experimentally.  Indeed, flux-gradient driven
vortex rivers similar to the ones described in this work have already
been imaged using Lorentz microscopy \cite{16}.

\section{Comparison with other avalanche studies}

Superconductors represent only one of the many systems exhibiting
avalanche behavior that have recently been studied.  In this section,
we briefly compare our work with a small sample of studies on avalanches
in other systems, focusing on dynamical instabilities in dissipative
extended systems which are very slowly driven {\it towards} (and
not away from) marginally stable states.  The literature on this
subject is vast, and it is not the goal of this section to review
it.

The work presented in this paper differs from previous studies
in several ways.  First, most theoretical studies on
avalanches (e.g., Ref. \cite{14},\cite{39}, and references therein) 
employ simple dynamical rules.  Our system evolves according to 
realistic equations of motion and uses a realistic range of physical
parameters.  Second, instead of using {\it ad hoc} dynamical
rules acting on discrete space and time variables, our simulations employ
a dynamics which is {\it continuous} in both space and time.
Third, most discrete dynamical rules proposed so far involve interactions
among {\it nearest-neighbor} cells, or at most next-nearest neighbor cells.
Every one of our vortices can interact with over 100 neighboring vortices,
making possible truly cooperative ``cascades'' in the 
marginally stable Bean critical state.  This type of highly
correlated motion is difficult to realistically model with
discrete maps.

Although most theoretical studies of avalanches have revealed no internal
structure in the dynamics of the avalanches, Ref. \cite{39} describes
a new series of extremal (i.e., uniformly driven) models that produce
avalanches with an interesting hierarchical structure of subavalanches
within avalanches, not found in the earlier simpler models.  These
subavalanches appear to be similar to the ones observed in our
vortex avalanches in the form of velocity oscillations, as shown
in Fig.~\ref{fig:2}.

Avalanches exhibiting similar broad distributions of sizes and lifetimes have
been observed experimentally in a variety of systems that are otherwise very
different.  In these systems, the movable objects can be vortices,
grains, electrons, or water droplets.  These interact through different
types of forces, from very short range (hard-core) interactions for
granular assemblies to longer range forces for electrons and vortices. 
The movable objects are driven by a variety of different forces (e.g.,
flux-gradient, Lorentz force, electrical current, gravity) and
dissipate energy while they are driven (e.g., due to particle-particle
collisions).  These systems can exhibit both static and dynamic
friction.  For instance, the pinned state of vortices is the
analog of static friction for grains, while the dissipative flux-flow
regime is the analog of the dynamic friction seen in vortices.  The
inertial effects are very important in some cases, such as granular
motion, and negligible in other cases, such as vortex motion.
The disorder in the sample can be frozen in, as in the quenched 
disorder produced by pinning sites in a superconductor, or dynamically
evolving, as in avalanching granular assemblies.  In spite of all
these differences, each of these systems exhibits avalanches with
broad distributions of sizes and lifetimes.  Some of the distributions follow
a power law over a limited range of values, resulting in
similar exponents (as seen in Table II).  Most
non-superconducting systems produce exponents in the range from 2 to 2.5.
Superconducting vortex avalanches can produce exponents that are
significantly below 2.  Indeed, the large variety of pinning landscapes
present in superconducting samples allows the possibility of observing
a range of exponents from 1.5 to 2.2.

Several words of caution are needed when comparing the exponents
presented in Table II.  First, the ranges over which power laws have been
observed in these systems are small, typically covering between one and three
orders of magnitude in the independent variable ($x$ axis).  
It would be ideal to probe the response of each
system over many more decades, but this is difficult to achieve 
experimentally.  Even with realistic numerical simulations it is difficult
to study a very large number of particles over a long period of time
in order to probe large avalanche sizes and lifetimes.  
Second, the quantities displayed in Table II are not identical.
In some cases, the measured ``avalanche size'' is the number of
particles (e.g., grains, vortices, droplets) falling off the edge of the
sample, while in other cases it refers to the actual number of
particles that moved in the sample without leaving it.  Third, the
manner and rate at which particles are added to the 

\begin{table}
\begin{tabular} {c|c|c}\hline
{Ref.} & {Method} & {Power law exponent} \\ \hline
{[9]} & {Water droplets} & 1.9 \\
{[7]} & {Magnetic garnet films} & 2.5 \\
{[1]} & {Granular pile} & 2.5 \\
{[42]} & {Granular pile} & 2.5 \\
{[2]} & {Granular pile} & 2.2 \\
{[6]} & {Rice grains} & 2.0 \\
{[5]} & {Ramped granular slides} & 2.1 \\
{[11]} & {Superconducting vortices} & {1.4 -- 2.2} \\
{[13]} & {Superconducting vortices} & {1.7 -- 2.2} \\ \hline
\end{tabular}
\caption{
Exponents of power laws observed in various experiments.
[9] considered avalanches in a continuous medium.  
[7] observed magnetic domains.  
[1] worked with ${\rm Al}_{2}{\rm O}_{3}$
particles and beach sand.  [42] used 3 mm iron or glass beads as
well as 1 -- 2 mm plastic beads.  [2] placed 0.4 -- 0.8 mm 
${\rm SiO}_{2}$ sand in piles of varying sizes.  [6] monitored a
quasi-one-dimensional pile of rice.  [5] imaged the surface behavior of 
sand avalanches.  [11] and [13] studied avalanches in ``hard''
(i.e., with srong pinning sites) superconductors.}
\end{table}

\hspace{-13pt}
system can affect
the values of the power law exponents or even alter the functional
form of the avalanche distributions.  In some cases, 
incoming particles are added in the bulk, while in others, particles
are added only at one edge.  For instance, experiments in 
which
incoming particles 
are randomly sprinkled on the sample are driven
in the bulk, while in experiments such as Ref.~\cite{6} and \cite{11},
the particles are added on one edge only, resulting in different
exponents.  Simulations also produce different exponents for
bulk driving \cite{39} and for driving on one edge \cite{40,41}.
Fourth, comparisons among systems displaying avalanche dynamics are difficult
to make because of the significant differences among these complex
systems, as noted above.

Many questions still remain open, including why so many different systems
exhibit broad distributions of avalanche sizes and lifetimes, what the
origin of the power law behavior is, and why the values of the power law
exponents range between 1.5 and 2.5.  In spite of considerable efforts,
a complete and convincing answer to these and other related questions
is still lacking.

\section{Summary}

We have quantitatively shown how the microscopic pinning parameters 
determine the nature of the avalanches that occur in superconducting
samples driven very slowly by an increasing external magnetic field.  
By using large-scale MD simulations to monitor
the vortices participating in avalanches,
we observe motion along winding paths through the sample,
and find that each vortex moves only one to two pinning sites 
during an avalanche rather than crossing the entire sample.  
Most avalanches are small and are contained
inside the sample.  Thus, they cannot be
detected with experiments that probe only vortices exiting the sample.

Avalanches produce two kinds of vortex motion: pin to pin, and
extremely small displacements inside the small parabolic wells.
Only vortices that move from pin to pin directly participate in the
avalanches, suddenly releasing accumulated stress in the vortex
lattice through a succession of choppy bursts.
The vortices that do not directly participate in a given avalanche experience
extremely small displacements (typically $d_{i} \ll d_{p}$).
These small shifts help to slowly build up and transmit stress
throughout the vortex lattice.

Pinning strength affects the
average lifetime of a given avalanche by determining under how much
stress the lattice is held.
A high density of strong pinning holds the vortex lattice under a great deal
of accumulated stress, localizes avalanche disturbances, and is effective
at retaining vortices inside the sample.  
The resulting avalanches are suppressed in width, involve rapidly 
moving vortices, and have very short lifetimes.
Thus, avalanches in samples with strong dense pinning are best characterized
by plastic transport that occurs in brief, choppy bursts along
narrow vortex paths \cite{25}.
Weak or sparse pinning produces
little build up of stress in the vortex lattice.  
The avalanche events in such samples
involve vortices moving slowly in long-lasting events along much
broader vortex channels.
Higher vortex mobility in samples with weak pinning leads to larger total
vortex displacement $d_{\rm av}$ \cite{25}.  

The presence or absence of distinct channels for flow 
leads to a crossover from broad distributions to characteristic 
avalanche sizes.
At low pin densities, avalanches pulse
through the sample in narrow heavily trafficked
winding channels composed of interstitially pinned vortices.
As pin density increases, pin-to-pin vortex motion dominates,
the isolated channels disappear, 
and avalanches 
display a broad distribution over more
than a decade. 
An important, and non-intuitive, result of our simulations is that
the size of the critical current is not a good indicator of
broad distributions.  
Indeed, samples with {\it very} different critical currents $J_{c}$ (e.g., 
a sample with $f_{p}^{\rm max}=3.0f_{0}$, $n_{p}=0.96/\lambda^{2}$,
and $J_{c} \sim 0.03 \Phi_{0}c/2\pi$
and a second sample with $f_{p}^{\rm max}=0.3f_{0}$,
$n_{p}=5.93/\lambda^{2}$, and $J_{c} \sim 0.01 \Phi_{0}c/2\pi$) 
may have very different breadths to
their distributions;  the presence of unique channels, not the critical
current, is the important factor.

We note that avalanches in the Bean state cannot
be characterized by universal avalanche distributions valid
for all values of the pinning parameters.
Although the Bean state is always critical, it 
does not always display avalanches with a lack of characteristic scale,
and in those cases where avalanche distributions take the form of
a power law, the exponent varies with pinning \cite{11,13}.
Our results are consistent with experiments and explain the
sample- and  regime-dependence of recent experiments\cite{11,13}.

The authors acknowledge very 
helpful discussions with S. Field and J. Groth.
Computer services were provided by the Maui High Performance Computing Center,
sponsored in part by the Phillips Laboratory,
Air Force Materiel Command, USAF, under cooperative agreement
number F29601-93-2-0001.  Computing services were also provided by 
the University of Michigan Center for Parallel Computing, 
partially funded by NSF grant CDA-92-14296.
C.O. acknowledges support from the NASA Graduate Student Research Program.


\begin{references}
\bibitem{24} This relatively long interaction range is difficult to 
realistically simulate using cellular automata, but is very 
important for the production of ``domino'' effect cascades in the 
critical state, since it allows each vortex to interact
with up to 100 neighbors.
\end{references}
\end{document}